\documentclass[twocolumn]{aastex701}

\usepackage{booktabs}
\usepackage{multirow}
\usepackage{threeparttable}

\begin{document}

\title{Unsupervised Chemo-Dynamical Dissection of the Inner Galactic Halo: Discovery of Five Accreted Substructures with SDSS-V and Gaia}

\author[orcid=0000-0002-9993-7244,sname='Akbaba']{Furkan Akbaba}
\affiliation{Institute of Graduate Studies in Science, Istanbul University, Istanbul, Turkey}
\email[show]{furkan.akbaba@ogr.iu.edu.tr}  

\author[orcid=0000-0002-0435-4493, sname='Plevne']{Olcay Plevne} 
\affiliation{Faculty of Science, Department of Astronomy and Space Sciences, Istanbul University, Istanbul, Turkey}
\email{olcayplevne@istanbul.edu.tr}

\begin{abstract}
The inner Galactic halo is a complex graveyard of the Milky Way's earliest accretion events, where severe orbital phase-mixing challenges traditional dynamical stream-finding techniques. In this work, we present a purely data-driven, 12-dimensional chemo-dynamical analysis of the inner halo using the latest data from the \textsl{SDSS-V Milky Way Mapper} (DR19) and \textsl{Gaia} DR3. Utilizing an unsupervised machine learning framework based on UMAP and HDBSCAN, we perform a blind search for clustered populations within a chemically selected \textit{ex-situ} sample of 2,185 stars without imposing kinematic pre-selection criteria. Our pipeline successfully recovers nine kinematic groupings corresponding to seven previously known canonical substructures (including \textsl{Gaia}-Enceladus/Sausage, the Helmi Streams, and Sequoia), validating the robustness of the high-dimensional feature space. Furthermore, we report the discovery of five new tightly bound candidate substructures, designated FO1--FO5 ($E_{\rm tot} \leq -1.8 \times 10^5~\mathrm{km^2~s^{-2}}$). Statistical validation confirms four of the five candidates (FO1, FO3, FO4, and FO5) as robust chemo-dynamical overdensities, while the tentative candidate FO2 exhibits a striking nitrogen enhancement ($[\mathrm{N/Fe}] = +0.83 \pm 0.16$) indicative of the tidal debris from a disrupted massive globular cluster. Finally, we demonstrate that high-dimensional chemical information is critical for resolving dynamical degeneracies in the crowded inner halo, clearly differentiating structures that share similar orbits but distinct chemical signatures (e.g., FO5 and Shiva), as well as the reverse case where chemical similarity masks dynamical differences (e.g., FO3 and the Helmi Streams). These findings emphasize that the deepest regions of the Galactic potential remain far from fully mapped, preserving a rich and diverse record of the Galaxy's assembly history.
\end{abstract}



\section{Introduction}
\label{sec:intro}

The Milky Way has grown through numerous merger and accretion events over billions of years. As a result of these processes, the remnants of disrupted dwarf galaxies, globular clusters, and stellar streams can still be observed today as fossil structures within the Galactic stellar halo. Investigating the chemical and dynamical properties of these structures constitutes one of the primary goals of Galactic archaeology, which aims to reconstruct the formation history of the Galaxy.

The Galactic halo provides an environment capable of preserving the signatures of past merger events more efficiently than other Galactic components due to its long dynamical timescales and relatively low stellar density. Consequently, halo stars serve as a unique laboratory for studying the chemical and dynamical imprints of past accretion events \citep{Helmi2020, Horta2023, Dodd2025}.

In recent years, the combination of the precise astrometric measurements delivered by the \textsl{Gaia} mission \citep{Gaia16, GaiaDR3} with large-scale spectroscopic surveys such as APOGEE \citep{Apogee}, GALAH \citep{Galah}, LAMOST \citep{Zhao2012}, H3 \citep{Conroy2019}, RAVE \citep{Steinmetz2020}, and SEGUE \citep{Yanny2009} has revolutionized the field of Galactic archaeology. This wealth of data has enabled the discovery of numerous halo substructures, including \textsl{Gaia}-Enceladus/Sausage (GES; \citealt{Belokurov2018, Helmi2018, Haywood2018, Mackereth2019}), Sequoia \citep{Barba2019, Matsuno2019, Myeong2019}, Thamnos \citep{Koppelman2019}, LMS-1 \citep{Yuan2020}, Arjuna and I’itoi \citep{Naidu2020}, Heracles \citep{Horta2021}, and the recently identified Shiva and Shakti structures \citep{Malhan2024}. Detailed investigations of their chemical abundance patterns, age distributions, and dynamical properties \citep{Das2020, Hasselquist2021, Buder2022, Horta2023} have significantly improved our understanding of the Milky Way’s merger history.

Despite these advances, the inner Galactic halo still remains highly complex and difficult to disentangle. Due to strong phase mixing, stars originating from ancient merger events have become spatially dispersed over time. Furthermore, remnants of different accretion events substantially overlap with the Milky Way's \textit{in-situ} populations within integrals-of-motion space \citep{Belokurov2022, Rix2022}. This significantly increases the risk that selection methods relying solely on dynamical parameters may group physically unrelated structures into the same population.

Recent discoveries such as the Shiva and Shakti structures \citep{Malhan2024} have demonstrated that systems exhibiting similar orbital properties may nevertheless possess fundamentally different chemical origins. Consequently, modern Galactic archaeology increasingly requires data-driven approaches capable of simultaneously analysing high-dimensional chemical and dynamical information.

In this context, unsupervised machine learning techniques have become increasingly common in Galactic archaeology studies \citep{Yuan2018, Lovdal2022, Neitzel2025, Plevne2025}. In particular, high-dimensional approaches combining chemical and dynamical information help reduce the limitations of classical selection methods based on only a few parameters or two-dimensional projections, enabling the identification of more complex substructures. Methods such as Principal Component Analysis (PCA), Uniform Manifold Approximation and Projection (UMAP), and HDBSCAN provide powerful tools for identifying compact and chemically coherent stellar groups in high-dimensional chemo-dynamical space.

In this work, we combine \textsl{SDSS-V Milky Way Mapper} DR19 and \textsl{Gaia} DR3 data to investigate hidden accreted substructures within the inner halo of the Milky Way. After applying a PCA-based pre-separation in chemical space to isolate the ex-situ population, we search for new substructure candidates using UMAP and HDBSCAN clustering within a 12-dimensional chemo-dynamical feature space.

This paper is structured as follows. In Section~\ref{sec:data}, we describe the target selection criteria and the parent sample derived from the \textsl{SDSS-V} MWM DR19 and \textsl{Gaia} DR3 datasets. Section~\ref{sec:methods} details our unsupervised machine learning pipeline, including the chemical isolation via PCA and the subsequent clustering in the 12-dimensional chemo-dynamical parameter space using UMAP and HDBSCAN. In Section~\ref{sec:results}, we present our results, detailing the recovery of canonical literature structures and the properties of the five new candidate substructures. Section~\ref{sec:discussion} provides a comprehensive physical interpretation of these groups, discussing their potential origins as either disrupted star clusters or remnants of ancient accretion events. Finally, we summarize our core findings and conclude in Section~\ref{sec:conclusions}.

\section{DATA AND SAMPLE SELECTION} \label{sec:data}

\subsection{Spectroscopic and Astrometric Data}     

\noindent This study relies on combining astrometric data from the \textsl{Gaia} Data Release 3 \citep[DR3;][]{GaiaDR3} with spectroscopic observations from the nineteenth data release (DR19) of the \textsl{SDSS-V Milky Way Mapper} \citep[\textsl{MWM};][]{Kollmeier2026}. As a core component of the fifth-generation Sloan Digital Sky Survey \citep[\textsl{SDSS-V}:][]{Kollmeier2017}, the \textsl{MWM} provides high-resolution ($R\approx22,500$) near-infrared spectra across both hemispheres. This full-sky coverage is achieved through twin spectrographs \citep{Wilson2019} installed on the 2.5m Apache Point Observatory telescope in New Mexico \citep{Gunnetal2006} and the Du Pont telescope at Las Campanas Observatory in Chile \citep{BowenVaughan1973}. For the DR19 dataset, stellar parameters and elemental abundances for $\approx1,400,000$ stars were derived using the newly developed \texttt{astra} framework (A. R. Casey et al., in prep). This framework implements a modified \texttt{ASPCAP} pipeline \citep{Garcia2016}, powered by the \texttt{FEREE} code \citep{Allende2006, Allende2015} alongside atomic and molecular line lists from \citet{Hasselquist2016}, \citet{cunha2017}, and \citet{Smith2021}. Comprehensive details regarding the \textsl{MWM} DR19 target selection and the \texttt{ASPCAP} parameter extraction are described in \citet{Kollmeier2026} and \citet{MWM19}, respectively.

The \textsl{Gaia} space observatory \citep{Gaia16} has provided precise astrometry---including positions, proper motions, and parallaxes---alongside low-resolution optical spectroscopy for $\approx1.5$ billion sources down to an apparent magnitude limit of $G < 20.7$. In this work, we utilize the astrometric solutions provided in \textsl{Gaia} DR3 \citep{GaiaDR3}, complemented by the photo-geometric distance estimates derived by \citet{Bailer21}.

By merging the radial velocities obtained from \textsl{MWM} with \textsl{Gaia}'s astrometry, we construct a complete 6D phase-space catalog required to compute stellar kinematics and orbits. To achieve this, we map the observational data (radial velocities, proper motions, and sky coordinates) into a Galactocentric Cartesian frame utilizing the \texttt{Astropy} library \citep{Astropy2013, Astropy2018, Astropy2022}. Our adopted coordinate framework assumes a solar distance to the Galactic center of $R_{\odot} = 8.275$ kpc \citep{Gravity2021}, a height above the midplane of $z_{\odot} = 0.02$ kpc \citep{Bennett2019}, and a local standard of rest (LSR) velocity of $V_{\mathrm{LSR}} = 232.8$ km s$^{-1}$. The Solar peculiar velocity relative to the LSR is set to $(U, V, W) = (11.10, 12.24, 7.25)$ km s$^{-1}$ \citep{Schonrich2010}. Finally, we integrated the stellar orbital trajectories using the \texttt{galpy} \texttt{python} package \citep{galpy}, adopting the \citet{McMillan2017} model for the Galactic gravitational potential.
\subsection{Sample Selection}

Starting from the full \textsl{MWM} DR19 catalogue containing approximately $1.4\times10^{6}$ stars, we applied a series of quality and selection criteria to construct the final sample used throughout this work.

\begin{description}

\item[Red Giant Branch stars] 
We restricted the sample to red giant branch (RGB) stars using the \textsl{MWM}-derived atmospheric parameters, selecting stars with effective temperatures and surface gravities satisfying $3{,}500 \leq T_{\mathrm{eff}} \leq 5{,}500~\mathrm{K}$ and $0 < \log g < 3.6$, respectively.

\item[Medium signal-to-noise spectra] 
To ensure robust spectroscopic measurements, only stars with spectral signal-to-noise ratios satisfying $\mathrm{S/N} > 50$ were retained.

\item[High-quality derived spectral parameters] 
We retained only stars with \texttt{pipeline\_flags} = 0 and \texttt{spectrum\_flags} = 0. In addition, reliable measurements were required for the atmospheric parameters, radial velocities, and all chemical abundance dimensions used throughout the PCA analysis.

\item[High-quality distance estimates] 
We adopted the \textsl{Gaia} photo-geometric distance estimates of \citet{Bailer21} and required distance uncertainties smaller than $20\%$, corresponding to $d/\sigma_d > 5$. In addition, as discussed in Section~\ref{sec:correction}, we restricted the sample to stars located within 5~kpc from the Sun ($d_\odot < 5$~kpc).

\item[Removal of globular cluster members] 
Stars with globular cluster membership probabilities larger than 0.5 were excluded using the catalogue of \citet{Vasiliev2021}.

\item[Abundance-quality cuts] 
Since our analysis relies on chemical abundance patterns, we imposed abundance-quality constraints on the six chemical dimensions used throughout the PCA analysis: [Fe/H], [Mg/Fe], [Al/Fe], [Mn/Fe], [Ni/Fe], and [Si/Fe]. For each quantity, the corresponding abundance quality flag was required to be equal to zero.

\item[Reduction of disk contamination] 
Because the parent sample is strongly dominated by disk populations, their overwhelming contribution may bias the machine-learning analysis and dilute chemically distinct accreted structures. To alleviate this effect, we followed \citet{Fernandes2023} and restricted the sample to stars with orbital eccentricities satisfying $e > 0.25$.

\end{description}

After applying all quality cuts and selection criteria, the final parent sample used for the PCA and clustering analysis consists of 41,928 stars.

\subsection{Mitigating \texorpdfstring{$\log~g$}{log g}-driven Systematics in Chemical Abundances}
\label{sec:correction}

\begin{figure}
    \centering
    \includegraphics[width=\columnwidth]{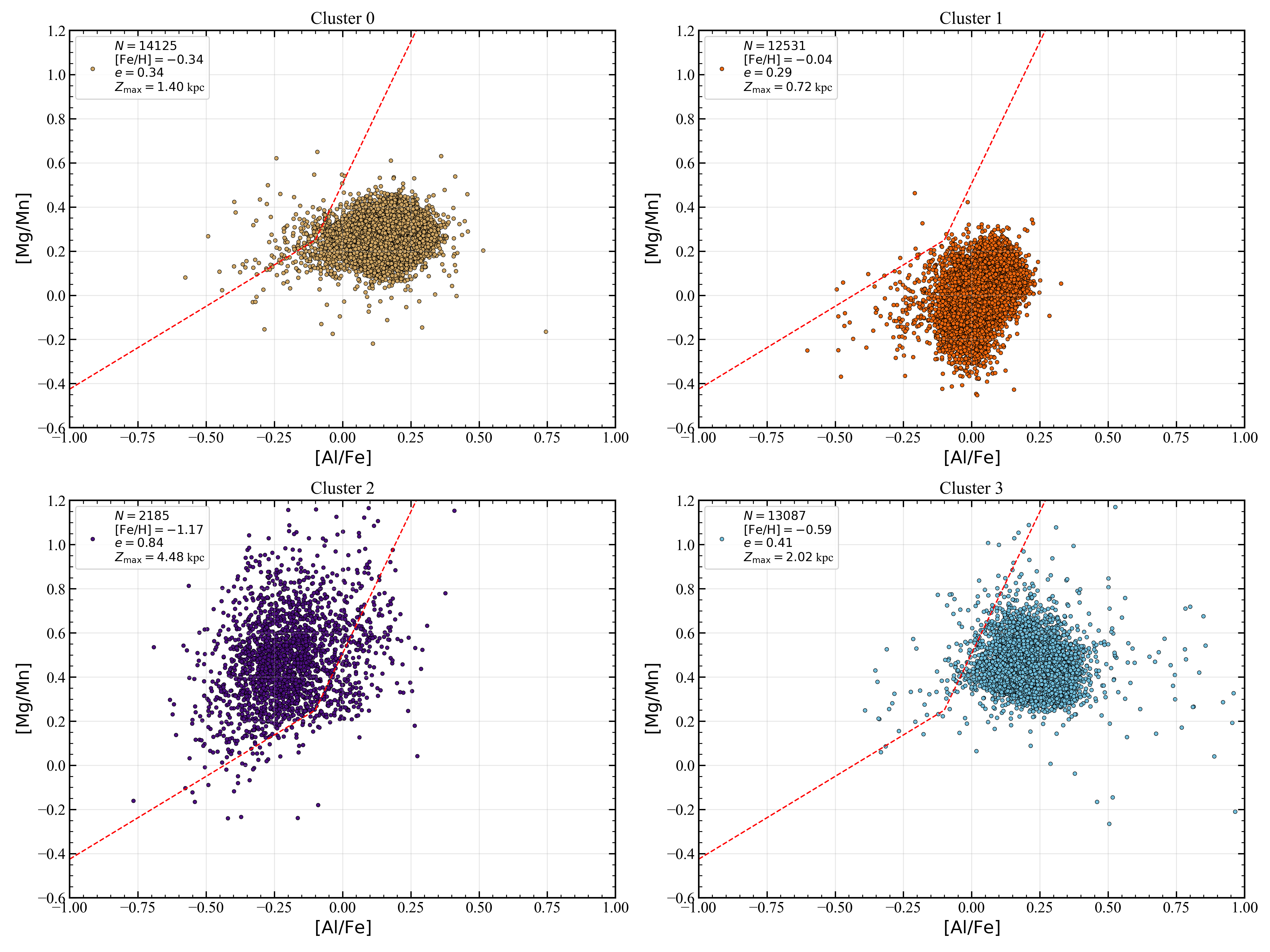}
    \caption{
Distribution of the four PCA+KMeans populations in the [Al/Fe]--[Mg/Mn] chemical diagnostic plane. Each panel corresponds to one of the four recovered PCA+KMeans clusters. The dashed curve indicates the empirical accreted/in-situ chemical boundary proposed by \citet{Alinder2025}. Cluster 2 (purple) lies predominantly within the chemically accreted region above the boundary, confirming its ex-situ nature, whereas the remaining clusters occupy the chemically in-situ region of the diagram.}
    \label{fig:al_fe_mg_mn}
\end{figure}

\noindent Extensive galactic surveys frequently exhibit correlations between derived elemental abundances and fundamental stellar parameters, particularly surface gravity ($\log~g$), as highlighted by several recent studies \citep{Eilers2022, Weinberg2022, Horta2023, Sit2024, Kisku2025}. These variations may stem either from genuine physical processes linked to stellar evolution or from pipeline-induced systematic offsets. To ensure robust and unbiased comparisons across stellar populations spanning diverse $\log~g$ ranges—which inherently vary with distance and survey magnitude limits—it is critical to calibrate out these parameter-dependent systematic effects. 

We implement the empirical calibration approach introduced by \citet{Horta2023}. Specifically, we model the $[\mathrm{X/H}]$ dependence on $\log~g$ using a second-order polynomial fit. The residual difference between this polynomial curve and the global median $[\mathrm{X/H}]$ value is evaluated across the $\log~g$ domain. By applying these residuals as an additive correction to the raw $[\mathrm{X/H}]$ measurements, we enforce a flattened, unbiased relationship between the corrected $[\mathrm{X/H}]$ abundances and $\log~g$. These adjusted values are subsequently used to compute our final, calibrated $[\mathrm{X/Fe}]$ ratios. The empirical effectiveness of this calibration procedure across the measured chemical dimensions is visually demonstrated and documented in Appendix~\ref{sec:app_correction} (see Figure~\ref{fig:abundance_correction}).

\begin{figure}
    \centering
    \includegraphics[width=\columnwidth]{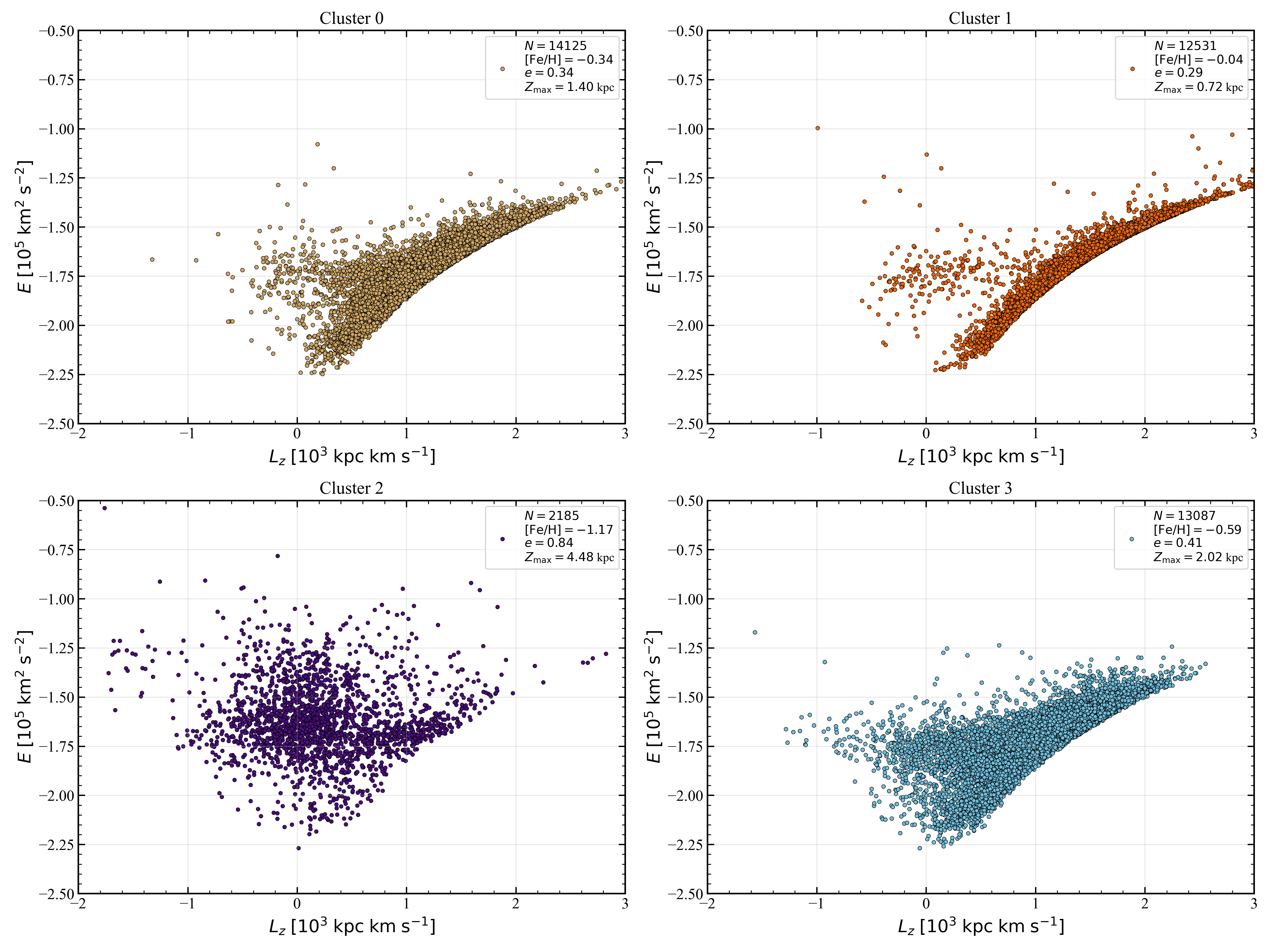}
    \caption{
Distribution of the four PCA+KMeans populations in the $E_{\rm tot}$--$L_z$ plane. Cluster 0 (tan; $N=14{,}125$), Cluster 1 (orange; $N=12{,}531$), and Cluster 3 (cyan; $N=13{,}087$) occupy dynamically colder regions associated with the Galactic disk populations, while Cluster 2 (purple; $N=2{,}185$) exhibits systematically higher orbital energies, a broader angular momentum distribution, and significantly larger orbital eccentricities and vertical excursions. The distinct dynamical configuration of Cluster 2 supports its interpretation as an accreted ex-situ population.}
    \label{fig:e_lz}
\end{figure}

As pointed out by \citet{Eilers2022}, distance-dependent selection biases naturally lead to a disproportionate fraction of lower-gravity (intrinsically brighter) stars at greater distances. Because the intrinsic chemical abundances of the Milky Way disk also exhibit spatial gradients across the Galaxy, a naive $\log~g$ correction applied over an unrestricted volume could inadvertently wash out authentic Galactocentric chemical signatures. To decouple these real spatial abundance gradients from the $\log~g$ systematics, we confine our calibration sample to a localized spatial footprint, restricting our selection to stars within a 5 kpc radius from the Sun. While imposing a strict $\log~g$ cut could theoretically circumvent this issue without spatial restrictions, such a stringent filter would severely deplete the dataset, thereby compromising the statistical power required for our subsequent clustering analysis.
\section{Methods} \label{sec:methods}

Our analysis proceeds in four main steps. First, we identify the ex-situ (accreted) stellar population using an unsupervised clustering analysis performed in chemical abundance space. Second, the resulting ex-situ sample is projected into a lower-dimensional manifold using UMAP to highlight potential substructures. Third, density-based clustering with HDBSCAN is applied to identify density overdensities within the ex-situ population. Finally, the recovered groups are compared with previously reported Galactic substructures from the literature.

\begin{figure}
    \centering
    \includegraphics[width=\columnwidth]{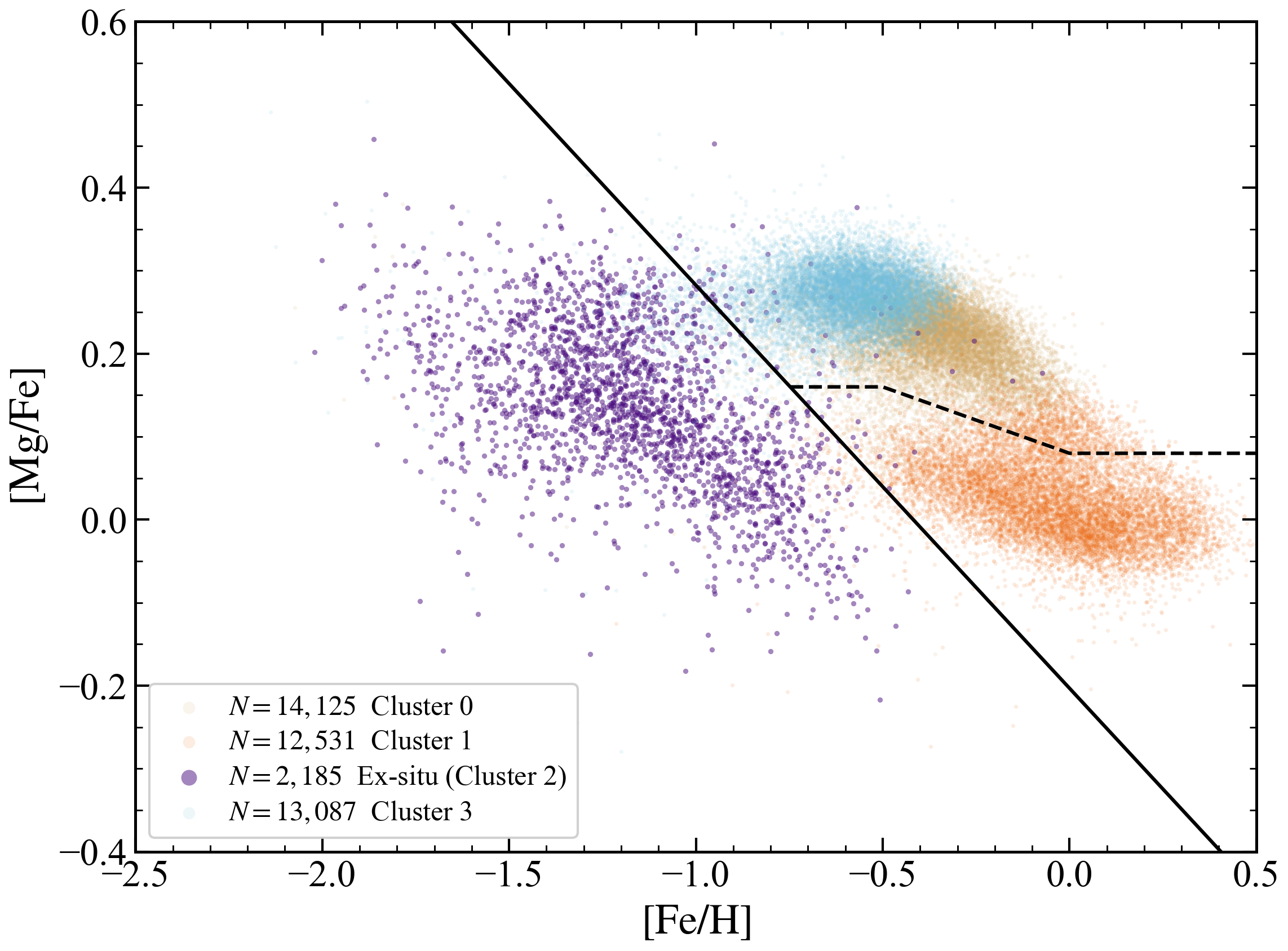}
    \caption{Distribution of the four PCA+KMeans populations in the [Mg/Fe]--[Fe/H] plane. The colour coding corresponds to the four chemically distinct populations identified through the PCA+KMeans analysis. Solid line is taken from \cite{Mackereth2019} and dashed line is taken from \cite{Akbaba2026}.}
    \label{fig:mg_fe_fe_h}
\end{figure}

\subsection{Identification of the Ex-Situ Population}

Chemical abundances preserve information about the nucleosynthetic history of the progenitor system and are generally less sensitive to phase mixing than purely dynamical quantities. For this reason, the initial separation between in-situ and ex-situ populations was performed in chemical abundance space. To characterise the chemical properties of the sample, we constructed a six-dimensional abundance vector for each star using the corrected abundance ratios:

\begin{equation}
F = \left(
[\mathrm{Fe/H}],
[\mathrm{Mg/Fe}],
[\mathrm{Al/Fe}],
[\mathrm{Mn/Fe}],
[\mathrm{Ni/Fe}],
[\mathrm{Si/Fe}]
\right)
\end{equation}

As described in Section~\ref{sec:data}, stars with orbital eccentricities satisfying $e > 0.25$ were selected in order to reduce the dominant contribution of dynamically cold disk populations. Prior to the clustering analysis, all chemical dimensions were standardised to zero mean and unit variance using a \texttt{StandardScaler}. This prevents dimensions with larger dynamic ranges from dominating the variance decomposition. We then applied Principal Component Analysis (PCA) to the standardised chemical abundance space and retained the first three principal components (PC1--PC3), which together explain approximately 91\% of the total variance within the chemical abundance space. The reduced three-dimensional PCA representation was subsequently clustered using the K-means algorithm with $k=4$ (\texttt{n\_init}=50, \texttt{random\_state}=13). The choice of $k=4$ was guided by the elbow criterion and by the stability of the resulting chemical populations. The PCA+KMeans analysis revealed four chemically distinct stellar populations. To identify the accreted component among these groups, we examined the clusters in the [Al/Fe]--[Mg/Mn] plane following the chemical separation method introduced by \citet{Horta2021}. The distribution of the four PCA+KMeans populations in this plane is shown in Figure~\ref{fig:al_fe_mg_mn}.

\begin{figure*}
    \centering
    \includegraphics[width=\textwidth]{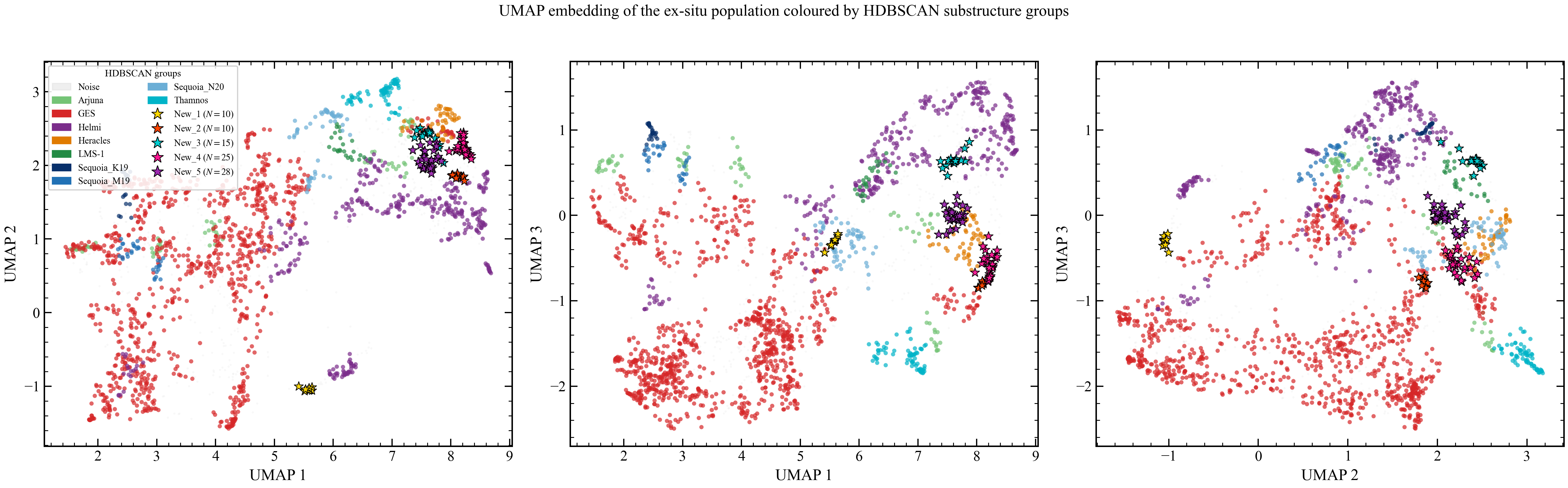}
    \caption{
Three pair-wise projections of the three-dimensional UMAP embedding constructed from the 12-dimensional chemo-dynamical feature space of the ex-situ sample ($N=2185$). The adopted feature space consists of six chemical dimensions ([Fe/H], [Mg/Fe], [Al/Fe], [Mn/Fe], [Ni/Fe], [Si/Fe]) and six dynamical dimensions ($E_{\rm tot}$, $L_z$, $e$, $Z_{\max}$, $J_R$, and $J_z$). Points are colour-coded according to their HDBSCAN cluster assignments.}
    \label{fig:umap_hdbscan}
\end{figure*}

The chemical nature of the selected ex-situ population is further validated in the [Mg/Fe]--[Fe/H] plane shown in Figure~\ref{fig:mg_fe_fe_h}, where the accreted candidates occupy the low-metallicity, low-$\alpha$ sequence characteristic of chemically accreted stars, consistent with the separation proposed by \citet{Mackereth2019}. Their dynamical consistency is illustrated in the $E$--$L_z$ plane in Figure~\ref{fig:e_lz}.

Based on this chemical classification, the final ex-situ sample consists of 2,185 stars. This chemically identified accreted population forms the basis of all subsequent substructure analyses.

\subsection{UMAP Dimensionality Reduction}

After identifying the ex-situ population, we searched for possible substructures within this sample using a combined chemical and dynamical feature space in order to simultaneously trace similarities in both nucleosynthetic history and orbital properties. The adopted feature matrix consists of six chemical and six dynamical
dimensions:

\begin{itemize}
\item Chemical abundances:
[Fe/H], [Mg/Fe], [Al/Fe], [Mn/Fe], [Ni/Fe], [Si/Fe]

\item Dynamical quantities:
$E_{\rm tot}$, $L_z$, $e$, $Z_{\max}$, $J_R$, and $J_z$
\end{itemize}

The chemical and dynamical blocks were standardised independently before being combined into a single 12-dimensional feature matrix. To visualise and separate potential substructures in this high-dimensional space, we applied Uniform Manifold Approximation and Projection (UMAP; \citealt{McInnes2018}). UMAP is particularly effective for preserving local neighbourhood structure while enabling the identification of compact overdensities in complex high-dimensional datasets.

The UMAP analysis was performed using the following hyperparameters:

\begin{itemize}
\item \texttt{n\_components = 3}
\item \texttt{n\_neighbors = 20}
\item \texttt{min\_dist = 0.0}
\item \texttt{metric = euclidean}
\item \texttt{random\_state = 42}
\end{itemize}

The adopted UMAP hyperparameters were selected to preserve local neighbourhood structure while maintaining compact overdensity separation within the low-dimensional embedding. In particular, \texttt{n\_neighbors = 20} provided a balance between fragmented local structure and excessive smoothing of cluster boundaries, while \texttt{min\_dist = 0.0} enabled compact cluster packing suitable for subsequent density-based clustering with HDBSCAN.

Stars with missing values in any of the adopted dimensions were excluded from this step in order to ensure a homogeneous feature space for the subsequent clustering analysis.

\subsection{Substructure Identification with HDBSCAN}

The three-dimensional UMAP embedding was analysed using the Hierarchical Density-Based Spatial Clustering of Applications with Noise (HDBSCAN; \citealt{Campello2013}; \citealt{McInnes2017}) algorithm.

HDBSCAN is well suited for identifying Galactic substructures because it does not require the number of clusters to be specified a priori and can naturally identify overdensities with irregular shapes and varying density
contrasts. In addition, the algorithm assigns noise labels to stars that do not belong to statistically significant overdensities, reducing the risk of artificial cluster assignments.

The HDBSCAN analysis was performed using the following hyperparameters:

\begin{itemize}
\item \texttt{min\_cluster\_size = 10}
\item \texttt{min\_samples = 1}
\item \texttt{cluster\_selection\_method = eom}
\end{itemize}

The adopted \texttt{min\_cluster\_size} value reflects a compromise between sensitivity to small overdensities and robustness against fragmentation caused by statistical noise. The choice of \texttt{min\_samples = 1} maximises sensitivity to low-density substructures while allowing HDBSCAN to retain diffuse halo overdensities that may otherwise be classified as noise. The resulting HDBSCAN labels were subsequently analysed in order to identify statistically significant overdensities within the ex-situ population.

\subsection{Cluster Identification and Validation}

The recovered HDBSCAN groups were compared with the accreted substructures reported by \citet{Horta2023}.
For each detected group, we computed the median chemical and dynamical properties and compared them with the published reference values.

The comparison utilised both chemical and dynamical reference properties reported by \citet{Horta2023}, including the chemical abundance dimensions ([Fe/H], [Mg/Fe], [Al/Fe], [Si/Fe], [Mn/Fe], [Ni/Fe],
[O/Fe], [Ca/Fe], and [Ti/Fe]), together with the dynamical quantities ($E_{\rm tot}$, $e$, $Z_{\max}$, and $L_z$).

The comparison was performed using a normalised chi-square distance defined as:

\begin{equation}
\chi^2(O, R, D) =
\frac{1}{N_D}
\sum_{d \in D}
\left(
\frac{O_d - R_d^\mu}{R_d^\sigma}
\right)^2
\end{equation}

where $O_d$ represents the observed median property of the detected group, $R_d^\mu$ and $R_d^\sigma$ correspond to the literature median and standard deviation for a given dimension $d$, and $N_D$ is the number of valid dimensions included in the comparison.

Chemical and dynamical comparisons were evaluated separately and then combined into a single metric:

\begin{equation}
\chi^2_{\rm combined} =
\frac{\chi^2_{\rm chem} + \chi^2_{\rm kin}}{2}
\end{equation}

Each detected group was assigned to the literature substructure yielding the minimum $\chi^2_{\rm combined}$ value, while groups without convincing matches were classified as candidate new substructures.

A threshold of $\chi^2_{\rm combined} \leq 2.0$ was adopted to identify statistically consistent matches with previously reported substructures. This threshold approximately corresponds to an average deviation of $\sim1.4\sigma$ across the adopted dimensions, providing a conservative matching criterion.

\begin{figure*}
    \centering
    \includegraphics[width=\textwidth]{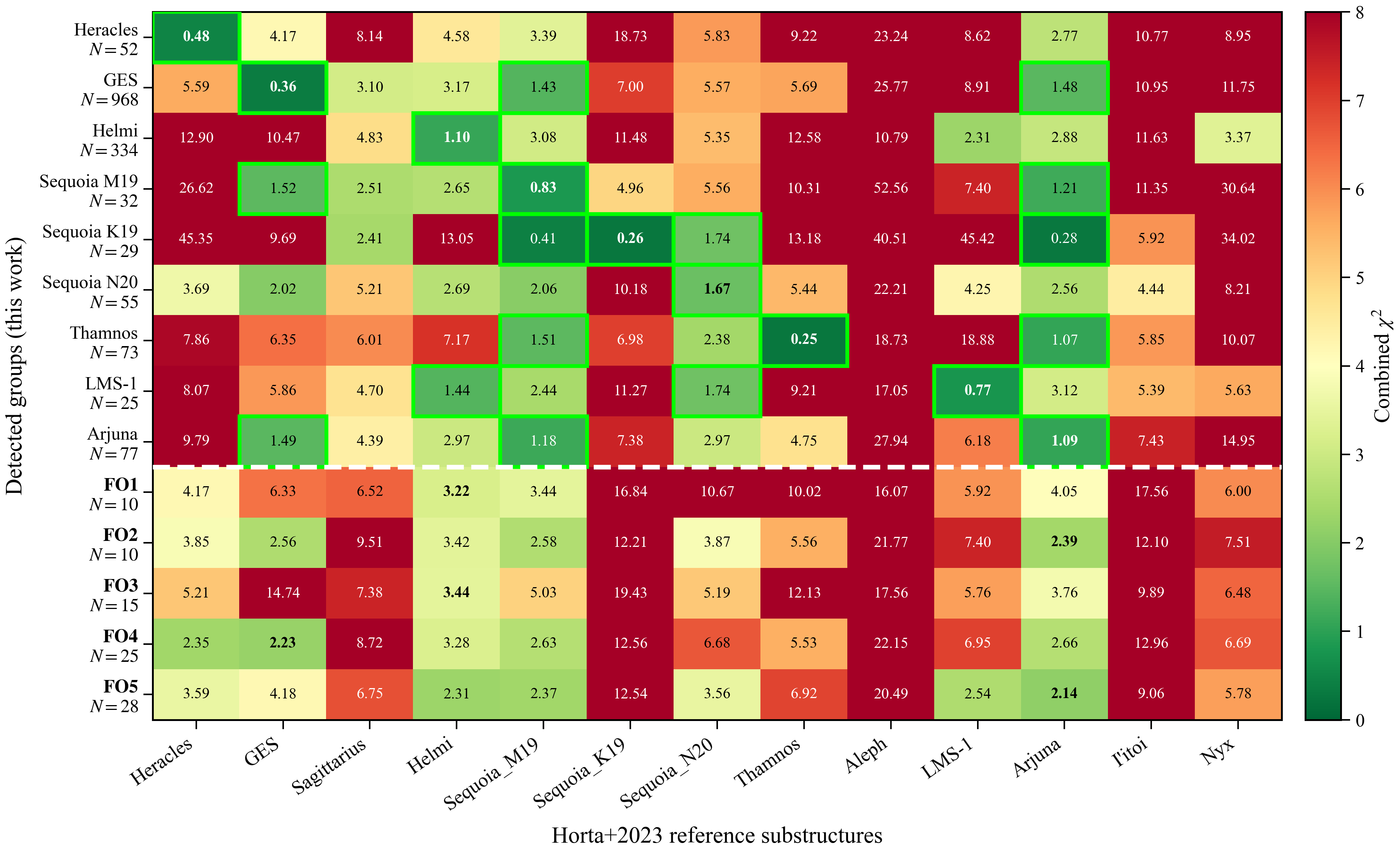}
    \caption{
Combined chemo-dynamical $\chi^2$ similarity matrix between the HDBSCAN groups identified in this work and the literature substructures compiled by \citet{Horta2023}. Rows correspond to the recovered groups and columns to the reference structures. The combined metric represents the average of the chemical and dynamical $\chi^2$ distances. Lower values indicate stronger similarity. Groups below the dashed line correspond to the newly identified candidate structures FO1--FO5.}
    \label{fig:ki_square}
\end{figure*}

Silhouette coefficients were computed in the three-dimensional UMAP embedding space in order to quantify the geometrical separation of each detected group. Positive silhouette values indicate that stars are, on average, more closely associated with their assigned group than with neighbouring overdensities, while negative values suggest possible overlap or unstable assignments.

In addition, a permutation-based cohesion test was performed in the full 12-dimensional feature space by comparing the within-group compactness of each candidate structure against random draws from the parent ex-situ population.

The robustness of the identified groups was further evaluated through bootstrap resampling tests. The bootstrap analysis consisted of repeating the full PCA$\rightarrow$KMeans$\rightarrow$UMAP$\rightarrow$HDBSCAN
pipeline over 30 bootstrap realisations using resampled datasets drawn with replacement in order to assess the stability of the recovered memberships.

Finally, the orbital distribution of the newly identified candidate groups was compared with the Shiva selection region recently proposed by \citet{Malhan2024} in the $E_{\rm tot}$--$L_z$ plane.
Stars falling within the published Shiva selection polygon were flagged in order to evaluate possible orbital overlap between the newly identified overdensities and previously reported inner-halo structures.

\section{Results}
\label{sec:results}

\begin{table*}
\caption{Mean $[X/\mathrm{Fe}]$ abundance values for each substructure identified in this work. (known structures)}
\label{tab:chem_known}
\resizebox{\textwidth}{!}{%
\begin{tabular}{lrrrrrrrrr}
\hline\hline
\multicolumn{1}{l}{} & \textbf{Arjuna} & \textbf{GES} & \textbf{Helmi} & \textbf{Heracles} & \textbf{LMS-1} & \textbf{Sequoia\_K19} & \textbf{Sequoia\_M19} & \textbf{Sequoia\_N20} & \textbf{Thamnos} \\
$N$ & 77 & 968 & 334 & 52 & 25 & 29 & 32 & 55 & 73 \\
\hline
{[C/Fe]} & $-0.32 \pm 0.36$ & $-0.28 \pm 0.27$ & $-0.17 \pm 0.30$ & $-0.28 \pm 0.18$ & $-0.26 \pm 0.69$ & $-0.29 \pm 0.38$ & $-0.31 \pm 0.25$ & $-0.19 \pm 0.47$ & $-0.21 \pm 0.28$ \\
{[N/Fe]} & $0.06 \pm 0.30$ & $0.01 \pm 0.28$ & $0.05 \pm 0.31$ & $0.22 \pm 0.22$ & $0.17 \pm 0.52$ & $0.10 \pm 0.34$ & $-0.03 \pm 0.27$ & $0.23 \pm 0.39$ & $0.11 \pm 0.29$ \\
{[O/Fe]} & $0.35 \pm 0.13$ & $0.28 \pm 0.14$ & $0.34 \pm 0.16$ & $0.33 \pm 0.10$ & $0.40 \pm 0.20$ & $0.24 \pm 0.12$ & $0.28 \pm 0.11$ & $0.36 \pm 0.25$ & $0.37 \pm 0.18$ \\
{[Na/Fe]} & $-0.07 \pm 0.59$ & $-0.17 \pm 0.56$ & $-0.17 \pm 0.59$ & $-0.29 \pm 0.49$ & $0.04 \pm 0.77$ & $-0.25 \pm 0.51$ & $-0.11 \pm 0.56$ & $-0.05 \pm 0.65$ & $0.02 \pm 0.60$ \\
{[Mg/Fe]} & $0.19 \pm 0.08$ & $0.10 \pm 0.09$ & $0.18 \pm 0.10$ & $0.22 \pm 0.05$ & $0.22 \pm 0.07$ & $0.08 \pm 0.06$ & $0.06 \pm 0.08$ & $0.21 \pm 0.11$ & $0.23 \pm 0.06$ \\
{[Al/Fe]} & $-0.22 \pm 0.12$ & $-0.23 \pm 0.13$ & $-0.14 \pm 0.17$ & $-0.11 \pm 0.09$ & $-0.37 \pm 0.13$ & $-0.34 \pm 0.15$ & $-0.27 \pm 0.11$ & $-0.36 \pm 0.12$ & $-0.15 \pm 0.12$ \\
{[Si/Fe]} & $0.22 \pm 0.06$ & $0.15 \pm 0.08$ & $0.20 \pm 0.09$ & $0.24 \pm 0.05$ & $0.21 \pm 0.07$ & $0.12 \pm 0.06$ & $0.15 \pm 0.06$ & $0.22 \pm 0.07$ & $0.25 \pm 0.05$ \\
{[Ca/Fe]} & $0.15 \pm 0.23$ & $0.13 \pm 0.15$ & $0.14 \pm 0.17$ & $0.17 \pm 0.10$ & $0.03 \pm 0.47$ & $0.08 \pm 0.20$ & $0.12 \pm 0.10$ & $0.01 \pm 0.44$ & $0.20 \pm 0.18$ \\
{[Ti/Fe]} & $-0.10 \pm 0.23$ & $-0.10 \pm 0.20$ & $-0.09 \pm 0.23$ & $-0.04 \pm 0.19$ & $-0.21 \pm 0.40$ & $-0.20 \pm 0.22$ & $-0.18 \pm 0.19$ & $-0.15 \pm 0.38$ & $-0.12 \pm 0.23$ \\
{[Cr/Fe]} & $-0.18 \pm 0.41$ & $-0.21 \pm 0.40$ & $-0.22 \pm 0.46$ & $-0.26 \pm 0.33$ & $-0.00 \pm 0.67$ & $-0.20 \pm 0.34$ & $-0.17 \pm 0.48$ & $-0.12 \pm 0.62$ & $-0.12 \pm 0.44$ \\
{[Mn/Fe]} & $-0.38 \pm 0.12$ & $-0.35 \pm 0.16$ & $-0.36 \pm 0.18$ & $-0.33 \pm 0.12$ & $-0.02 \pm 0.17$ & $-0.28 \pm 0.24$ & $-0.42 \pm 0.18$ & $-0.03 \pm 0.23$ & $-0.39 \pm 0.16$ \\
{[Co/Fe]} & $-0.24 \pm 0.44$ & $-0.24 \pm 0.46$ & $-0.21 \pm 0.47$ & $-0.18 \pm 0.35$ & $-0.22 \pm 0.54$ & $-0.33 \pm 0.52$ & $-0.36 \pm 0.50$ & $-0.09 \pm 0.65$ & $-0.23 \pm 0.50$ \\
{[Ni/Fe]} & $-0.07 \pm 0.09$ & $-0.11 \pm 0.09$ & $-0.08 \pm 0.09$ & $-0.03 \pm 0.06$ & $-0.07 \pm 0.10$ & $-0.10 \pm 0.07$ & $-0.12 \pm 0.07$ & $-0.05 \pm 0.12$ & $-0.06 \pm 0.09$ \\
{[Ce/Fe]} & $-0.01 \pm 0.48$ & $-0.05 \pm 0.45$ & $-0.05 \pm 0.47$ & $-0.03 \pm 0.30$ & $0.24 \pm 0.76$ & $-0.17 \pm 0.51$ & $-0.21 \pm 0.41$ & $-0.04 \pm 0.47$ & $-0.06 \pm 0.37$ \\
{[Fe/H]} & $-1.34 \pm 0.19$ & $-1.08 \pm 0.27$ & $-1.18 \pm 0.30$ & $-1.19 \pm 0.15$ & $-1.70 \pm 0.12$ & $-1.41 \pm 0.18$ & $-1.11 \pm 0.21$ & $-1.68 \pm 0.17$ & $-1.37 \pm 0.12$ \\
\hline
\end{tabular}
}
\end{table*}

We present the results of applying the four-step analysis pipeline described in Section~\ref{sec:methods} to the accreted stellar population selected from the \textsl{MWM} DR19 sample. Our methodology yields a labelled catalogue of 2,185 ex-situ candidate stars. Of these, 1,733 stars (79.3\%) are assigned to statistically significant chemo-dynamical substructures, while 452 stars (20.7\%) are classified as noise by the HDBSCAN algorithm. In total, we recover nine previously reported Galactic substructures
and identify five new candidate overdensities, designated as FO1--FO5, where FO refers to the initials of the lead authors.

\subsection{The Ex-Situ Subsample}

The initial PCA+K-means analysis identified a chemically distinct population occupying the low-[Al/Fe] and elevated-[Mg/Mn] region of chemical abundance space, consistent with the accreted locus defined by \citet{Horta2021}. This population also exhibits the highest median orbital eccentricity among the recovered PCA clusters ($e \approx 0.88$ compared to $e \approx 0.35$--$0.55$ for the remaining populations), confirming its association with the dynamically hot stellar halo.

The resulting ex-situ subsample contains 2,185 stars after requiring complete measurements in all chemo-dynamical dimensions used in the UMAP analysis. The distribution of these stars in the [Mg/Fe]--[Fe/H] plane is shown in Figure~\ref{fig:mg_fe_fe_h}, while their orbital distribution in the $E_{\rm tot}$--$L_z$ plane is shown in Figure~\ref{fig:e_lz}. In both parameter spaces, the selected stars occupy regions commonly associated with the Milky Way accreted halo, supporting the interpretation that the adopted chemical selection successfully isolates ex-situ populations.

\subsection{UMAP Embedding and Substructure Segregation}

The combined 12-dimensional chemo-dynamical feature matrix (six chemical and six dynamical dimensions) was projected into a three-dimensional UMAP embedding. The resulting embedding, colour-coded by HDBSCAN group membership, is shown in the three pair-wise projections of Figure~\ref{fig:umap_hdbscan}.

The embedding reveals a clear segregation of the major Galactic substructures. The \textsl{Gaia}-Enceladus/Sausage (GES) population, which dominates the sample ($N=968$), occupies a broad but coherent region of the embedding. In contrast, the Helmi Streams ($N=334$) form a distinct, well-separated prograde structure. The retrograde region of the embedding is populated by Thamnos and the Sequoia-related structures, consistent with their shared dynamical origin. Heracles and LMS-1 appear as compact overdensities concentrated toward the lowest-energy region of parameter space.

The three pair-wise projections demonstrate that the combined chemo-dynamical feature space preserves sufficient information to separate classical merger remnants without imposing explicit orbital selection boundaries. The five new candidate groups (FO1--FO5), highlighted with star symbols in Figure~\ref{fig:umap_hdbscan}, occupy regions that do not coincide with any of the recovered literature
substructures, qualitatively supporting their interpretation as distinct chemo-dynamical overdensities.

\subsection{Recovery of Known Galactic Substructures}

HDBSCAN identifies 74 individual density peaks within the three-dimensional UMAP space. After applying the merging procedure described in Section~\ref{sec:methods}, nine previously reported Galactic substructures are recovered with $\chi^2_{\rm combined} < 2.0$.

The similarity matrix between the recovered groups and the reference structures from \citet{Horta2023} is shown in Figure~\ref{fig:ki_square}. The majority of recovered structures exhibit minimum $\chi^2_{\rm combined}$ values well below the adopted threshold, while the FO groups remain systematically offset from the known substructure loci.

\begin{figure*}
    \centering
    \includegraphics[width=\textwidth]{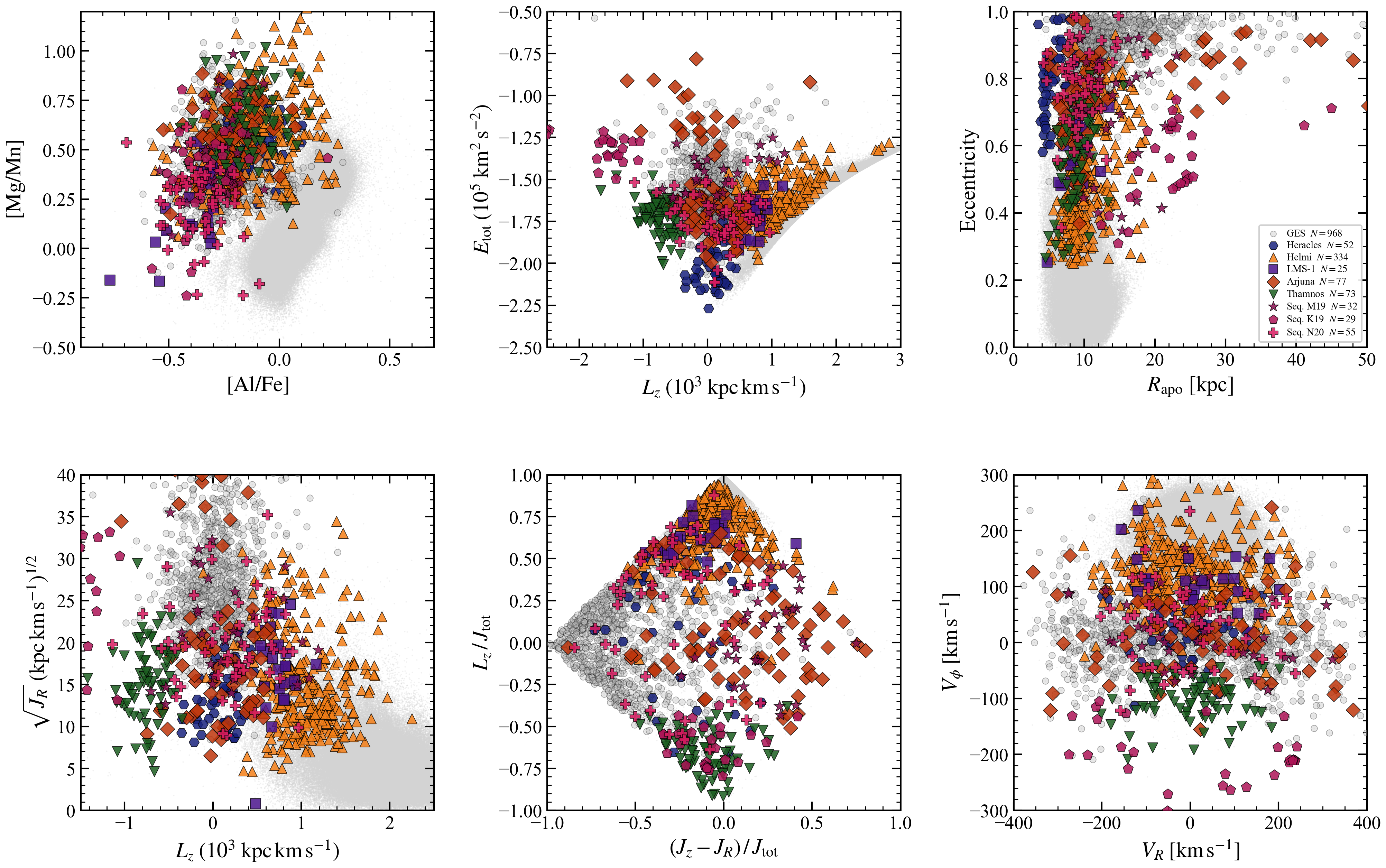}
    \caption{
Multi-dimensional overview of the recovered literature substructures. From left to right: [Mg/Mn]--[Al/Fe] plane, $E_{\rm tot}$--$L_z$ plane, eccentricity--$R_{\rm apo}$ plane, $\sqrt{J_R}$--$L_z$ action space, action diamond ($L_z/J_{\rm tot}$ versus $(J_z-J_R)/J_{\rm tot}$), and the $V_R$--$V_Z$ velocity plane. Grey points indicate the full ex-situ sample, while coloured symbols represent the recovered literature structures.}
    \label{fig:known_structures}
\end{figure*}

The mean chemical abundances and orbital parameters of the recovered structures are listed in
Table~\ref{tab:chem_known} and Table~\ref{tab:kin_known}, respectively. Their distributions across multiple chemo-dynamical diagnostic planes are shown in Figure~\ref{fig:known_structures}.

The dominant recovered component is GES ($N=968$, $\chi^2=0.36$), which spans a broad metallicity range
($[\mathrm{Fe/H}] = -1.08 \pm 0.27$) and exhibits the highest median orbital eccentricity
($e = 0.89 \pm 0.07$). The large spread in energy and angular momentum space reflects the well-known dynamical complexity of this major merger remnant.

The Helmi Streams constitute the second most populous recovered structure ($N=334$, $\chi^2=1.10$) and occupy the strongly prograde region of angular momentum space ($L_z = 1063~\mathrm{kpc~km~s^{-1}}$).
Their moderate eccentricities ($e = 0.48 \pm 0.15$) and relatively compact orbital distribution are clearly visible in Figure~\ref{fig:known_structures}.

The Sequoia family displays the expected retrograde orbital configuration. Among these groups, Sequoia\_K19 provides the most tightly constrained recovery ($\chi^2=0.26$, $N=29$),
with strongly retrograde angular momentum ($L_z = -1871~\mathrm{kpc~km~s^{-1}}$).
Thamnos is also recovered as a distinct retrograde population ($L_z = -757~\mathrm{kpc~km~s^{-1}}$),
while Sequoia\_N20 appears more tightly bound and chemically metal-poor than the other Sequoia variants.

Heracles ($N=52$) and LMS-1 ($N=25$) occupy the most tightly bound orbital regions among the recovered literature structures. Heracles exhibits the lowest median orbital energy ($E = -2.02\times10^5~\mathrm{km^2~s^{-2}}$), consistent with its proposed origin as a tightly bound
inner-halo progenitor. LMS-1 appears chemically metal-poor ($[\mathrm{Fe/H}] = -1.70$)
while maintaining moderate prograde angular momentum.

Collectively, the recovered structures preserve the expected chemical and orbital segregation patterns reported in previous studies, demonstrating that the adopted chemo-dynamical feature space successfully recovers both classical merger debris and more compact inner-halo overdensities.

While our pipeline successfully recovers the aforementioned canonical structures, four substructures from the \citet{Horta2023} reference catalogue — Sagittarius, Aleph, Nyx, and I'itoi — were not detected, each for distinct reasons rooted in our sample selection criteria. The Sagittarius stream \citep{Ibata1994} was entirely excluded by our $d < 5~\mathrm{kpc}$ distance limit, as its stellar population resides predominantly at heliocentric distances of $d \geq 10~\mathrm{kpc}$. Aleph and Nyx were removed at the halo pre-selection stage, as both structures have orbital eccentricities ($e_p \leq 0.25$) below our adopted threshold \citep{Naidu2020, Necib2020}. Finally, I'itoi \citep{Naidu2020} is represented by only $N=3$ reference stars in \citet{Horta2023}, which falls below our \texttt{min\_cluster\_size} $= 10$ parameter, rendering it statistically undetectable in the present sample. The non-recovery of these four structures therefore reflects well-defined selection boundaries rather than a limitation of the unsupervised methodology itself.

\subsection{New Candidate Substructures}

\begin{figure*}
    \centering
    \includegraphics[width=\textwidth]{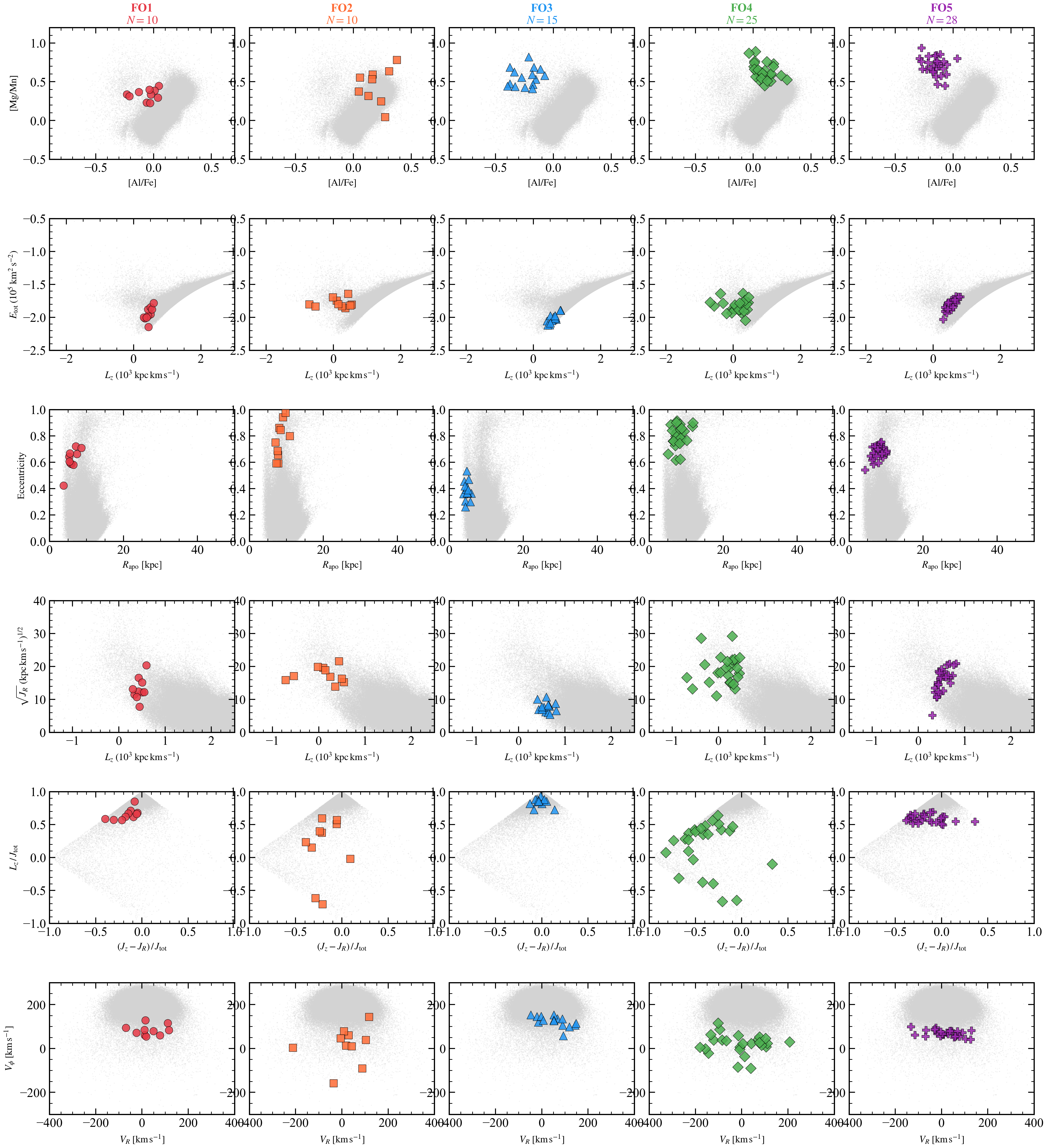}
    \caption{
Multi-dimensional overview of the five newly identified candidate structures FO1--FO5. Columns correspond to individual FO groups, while rows show different chemo-dynamical diagnostic spaces: [Mg/Mn]--[Al/Fe], $E_{\rm tot}$--$L_z$, eccentricity--$R_{\rm apo}$, $\sqrt{J_R}$--$L_z$, action diamond, and $V_R$--$V_Z$. Grey points indicate the full ex-situ population for reference.}
    \label{fig:fo_structures}
\end{figure*}

Five HDBSCAN groups lack satisfactory counterparts in the \citet{Horta2023} reference catalogue and are therefore classified as candidate new substructures. Additionally, we cross-matched the members of our five candidate groups with the catalogue of small local halo substructures recently identified by \citet{Dodd2025}. We found no overlapping stars between the two samples. Their chemical abundance properties and orbital parameters are summarised in Table~\ref{tab:chem_new} and Table~\ref{tab:kin_new}, respectively, while their distributions across multiple diagnostic planes are shown in Figure~\ref{fig:fo_structures}.

All five candidate groups occupy the chemically accreted region of the [Al/Fe]--[Mg/Mn] plane, yet display markedly different orbital configurations. Most notably, all groups reside at significantly more negative
orbital energies than the bulk of the known GES population, placing them among the most tightly bound populations of the inner halo.

\begin{table}
\caption{Mean $[X/\mathrm{Fe}]$ abundance values for each substructure identified in this work. (new candidate groups)}
\label{tab:chem_new}
\resizebox{\columnwidth}{!}{%
\begin{tabular}{lrrrrr}
\hline\hline
\multicolumn{1}{l}{} & \textbf{FO1} & \textbf{FO2} & \textbf{FO3} & \textbf{FO4} & \textbf{FO5} \\
$N$ & 10 & 10 & 15 & 25 & 28 \\
\hline
{[C/Fe]} & $-0.16 \pm 0.18$ & $-0.36 \pm 0.27$ & $-0.42 \pm 0.23$ & $-0.02 \pm 0.27$ & $-0.23 \pm 0.26$ \\
{[N/Fe]} & $0.07 \pm 0.17$ & $0.83 \pm 0.16$ & $0.23 \pm 0.23$ & $0.05 \pm 0.29$ & $0.06 \pm 0.33$ \\
{[O/Fe]} & $0.18 \pm 0.10$ & $0.16 \pm 0.14$ & $0.28 \pm 0.08$ & $0.39 \pm 0.10$ & $0.44 \pm 0.12$ \\
{[Na/Fe]} & $-0.31 \pm 0.54$ & $0.04 \pm 0.48$ & $-0.35 \pm 0.45$ & $-0.19 \pm 0.55$ & $-0.25 \pm 0.53$ \\
{[Mg/Fe]} & $0.10 \pm 0.03$ & $0.08 \pm 0.08$ & $0.21 \pm 0.04$ & $0.26 \pm 0.03$ & $0.24 \pm 0.04$ \\
{[Al/Fe]} & $-0.06 \pm 0.10$ & $0.28 \pm 0.28$ & $-0.24 \pm 0.10$ & $0.09 \pm 0.08$ & $-0.15 \pm 0.08$ \\
{[Si/Fe]} & $0.12 \pm 0.03$ & $0.19 \pm 0.05$ & $0.24 \pm 0.05$ & $0.27 \pm 0.03$ & $0.25 \pm 0.04$ \\
{[Ca/Fe]} & $0.12 \pm 0.07$ & $0.22 \pm 0.08$ & $0.15 \pm 0.08$ & $0.20 \pm 0.09$ & $0.24 \pm 0.08$ \\
{[Ti/Fe]} & $-0.08 \pm 0.11$ & $-0.03 \pm 0.17$ & $-0.11 \pm 0.09$ & $0.03 \pm 0.12$ & $-0.16 \pm 0.21$ \\
{[Cr/Fe]} & $-0.09 \pm 0.17$ & $-0.16 \pm 0.41$ & $-0.19 \pm 0.36$ & $-0.23 \pm 0.47$ & $-0.28 \pm 0.47$ \\
{[Mn/Fe]} & $-0.24 \pm 0.06$ & $-0.38 \pm 0.22$ & $-0.35 \pm 0.10$ & $-0.38 \pm 0.11$ & $-0.46 \pm 0.13$ \\
{[Co/Fe]} & $-0.13 \pm 0.25$ & $-0.58 \pm 0.38$ & $-0.23 \pm 0.42$ & $-0.11 \pm 0.56$ & $-0.21 \pm 0.49$ \\
{[Ni/Fe]} & $-0.07 \pm 0.03$ & $-0.11 \pm 0.10$ & $-0.02 \pm 0.04$ & $-0.10 \pm 0.06$ & $-0.03 \pm 0.06$ \\
{[Ce/Fe]} & $-0.08 \pm 0.50$ & $0.06 \pm 0.38$ & $-0.02 \pm 0.14$ & $0.03 \pm 0.50$ & $-0.22 \pm 0.40$ \\
{[Fe/H]} & $-0.83 \pm 0.08$ & $-1.39 \pm 0.13$ & $-1.34 \pm 0.11$ & $-1.06 \pm 0.16$ & $-1.33 \pm 0.11$ \\
\hline
\end{tabular}
}
\end{table}

\subsubsection{FO1 and FO3: Tightly Bound Prograde Structures}

FO1 contains 10 stars and represents the most metal-rich candidate group ($[\mathrm{Fe/H}] = -0.83 \pm 0.08$). It follows a tightly bound, low-$Z_{\rm max}$ orbit ($Z_{\rm max} = 1.83~\mathrm{kpc}$)
with moderate prograde angular momentum ($L_z = +447~\mathrm{kpc~km~s^{-1}}$).
Its compact distribution in the $E_{\rm tot}$--$L_z$ plane is clearly visible in Figure~\ref{fig:fo_structures}.

FO3 ($N=15$) is dynamically distinct from the remaining candidate groups due to its unusually circular orbit
($e = 0.38 \pm 0.07$), combined with low radial action ($J_R = 60~\mathrm{kpc~km~s^{-1}}$)
and small maximum orbital height ($Z_{\rm max} = 1.12~\mathrm{kpc}$). Its nearly planar orbit and enhanced $\alpha$-element abundances suggest a chemically accreted population with a dynamically cold orbital history. Such low-eccentricity configurations are comparatively uncommon among classical accreted halo structures.

\subsubsection{FO4 and FO5: High-Confidence Candidates}

FO4 ($N=25$) occupies a tightly bound region of the $E_{\rm tot}$--$L_z$ plane while exhibiting enhanced
$\alpha$-element abundances ($[\mathrm{Mg/Fe}] = 0.26 \pm 0.03$, $[\mathrm{Si/Fe}] = 0.27 \pm 0.03$).
Although its closest literature analogue is GES, its significantly more negative orbital energy
($E \approx -1.83\times10^5~\mathrm{km^2~s^{-2}}$) distinguishes it from the main GES population.

FO5 displays one of the most chemically compact abundance distributions among the candidate groups. It features elevated $\alpha$-element abundances, low manganese content ($[\mathrm{Mn/Fe}] = -0.46 \pm 0.13$), and a compact orbital distribution in action space. Collectively, these properties strongly support the interpretation of FO5 as a physically distinct accreted population. Interestingly, FO5 occupies a similar region of the $E_{\rm tot}$--$L_z$ plane as the recently identified Shiva substructure \citep{Malhan2024}, with approximately 39\% of its members falling within the Shiva selection polygon. However, the chemically accreted nature of FO5---particularly its low [Al/Fe] abundances---contrasts with the more [Al/Fe]-enhanced population typically associated with Shiva, suggesting a potentially distinct origin. A more detailed comparison and the implications of this overlap are discussed in Section~\ref{sec:discussion}.

\subsubsection{FO2: A Tentative Detection}

FO2 ($N=10$) is the weakest of the candidate groups. Although chemically consistent with an accreted population, it fails several stability tests discussed below. The group exhibits a negative silhouette score and weak bootstrap recovery fraction, suggesting that it may represent a local density fluctuation rather than a robust physical substructure.

\subsection{Validation and Stability Analysis}

The robustness of the candidate groups was evaluated using the four-tier validation framework described in
Section~\ref{sec:methods}. The resulting stability metrics are summarised in Table~\ref{tab:validation}.

FO1 and FO3 exhibit the strongest stability across all validation metrics. Both groups retain high co-clustering fractions across multiple UMAP neighbourhood scales, indicating that their recovery is not strongly dependent on the local embedding topology.

The cohesion analysis further demonstrates that FO1, FO3, FO4, and FO5 are significantly more compact in the
full 12-dimensional chemo-dynamical feature space than expected from random draws of the parent ex-situ sample ($p < 0.001$ for all groups). FO5 exhibits the highest overall compactness, while FO1 achieves the highest silhouette score, indicating strong geometrical separation from neighbouring structures in the UMAP embedding.

Bootstrap resampling confirms the robustness of FO1 and FO3, which maintain co-clustering fractions of
82.7\% and 71.6\%, respectively. FO4 and FO5 display moderate bootstrap stability, likely reflecting partial overlap with neighbouring high-density regions in the UMAP embedding, particularly near the GES and Arjuna loci.

FO2 remains the least stable candidate in all validation tests. Its bootstrap co-clustering fraction drops to only 16.1\%, and the group dissolves entirely under larger UMAP neighbourhood scales.
Accordingly, FO2 is retained as a tentative candidate pending confirmation from future surveys and larger samples.

\section{Discussion}
\label{sec:discussion}

The analysis of the \textsl{MWM} DR19 sample has revealed five new chemo-dynamical candidate substructures (FO1--FO5). In this section, we interpret their origins by chemical abundance patterns (Tables~\ref{tab:chem_known} and \ref{tab:chem_new}), combining their orbital properties (Tables~\ref{tab:kin_known} and \ref{tab:kin_new}),  and clustering robustness (Table~\ref{tab:validation}).

\subsection{Chemo-Dynamical Distinctiveness in the \texorpdfstring{$\chi^2$}{chi-square} Space}
\label{subsec:disc_chisq}

The recovered literature structures broadly reproduce the chemical and dynamical properties reported by \citet{Horta2023}, supporting the reliability of the adopted chemo-dynamical framework. In particular, the GES, Helmi, Sequoia, and Thamnos populations recover the expected orbital configurations and abundance trends associated with their previously identified accretion origins.

Against this reference framework, the newly identified FO groups occupy systematically offset regions in the combined chemo-dynamical parameter space. The combined chemo-dynamical distances ($\chi^2_{\rm comb}$) presented in Figure~\ref{fig:ki_square} and Table~\ref{tab:validation} reveal that FO1 and FO3 exhibit the largest separations from the \citet{Horta2023} reference catalogue. With values of $\chi^2_{\rm comb} = 3.162$ and $3.416$, respectively, they remain well beyond the adopted association threshold ($\chi^2 < 2.0$), suggesting that they are unlikely to represent simple extensions of the known accreted structures.

In contrast, FO2 ($\chi^2_{\rm comb} = 2.138$), FO4 ($\chi^2_{\rm comb} = 2.171$), and FO5 ($\chi^2_{\rm comb} = 2.156$) reside closer to the reference loci of Arjuna and GES. However, their detailed chemical abundance patterns and validation metrics indicate that these similarities are only partial, supporting the interpretation that they may trace dynamically or chemically distinct accretion debris.

\begin{table*}
\caption{Mean orbital parameter values for each substructure identified in this work (McMillan 2017 potential). (known structures)}
\label{tab:kin_known}
\resizebox{\textwidth}{!}{%
\begin{tabular}{lrrrrrrrrr}
\hline\hline
\multicolumn{1}{l}{} & \textbf{Arjuna} & \textbf{GES} & \textbf{Helmi} & \textbf{Heracles} & \textbf{LMS-1} & \textbf{Sequoia\_K19} & \textbf{Sequoia\_M19} & \textbf{Sequoia\_N20} & \textbf{Thamnos} \\
$N$ & 77 & 968 & 334 & 52 & 25 & 29 & 32 & 55 & 73 \\
\hline
$v_R$ & $19.77 \pm 168.88$ & $3.47 \pm 187.08$ & $-3.72 \pm 104.65$ & $-13.80 \pm 79.11$ & $18.81 \pm 102.26$ & $12.70 \pm 186.99$ & $7.31 \pm 158.18$ & $-5.30 \pm 109.80$ & $6.93 \pm 90.14$ \\
$v_\phi$ & $7.30 \pm 70.05$ & $4.40 \pm 59.15$ & $135.51 \pm 54.80$ & $23.13 \pm 50.37$ & $106.74 \pm 47.80$ & $-209.88 \pm 68.17$ & $25.60 \pm 65.36$ & $30.27 \pm 65.34$ & $-103.37 \pm 37.69$ \\
$v_Z$ & $-5.21 \pm 158.61$ & $-1.28 \pm 73.91$ & $-1.39 \pm 73.76$ & $5.56 \pm 60.39$ & $-26.45 \pm 73.97$ & $40.32 \pm 162.84$ & $-23.01 \pm 186.51$ & $-8.09 \pm 70.17$ & $-5.66 \pm 70.77$ \\
$L_x$ & $9.18 \pm 285.30$ & $-2.76 \pm 135.91$ & $-91.40 \pm 267.99$ & $14.11 \pm 98.33$ & $-74.01 \pm 172.75$ & $123.75 \pm 563.53$ & $-38.27 \pm 349.44$ & $-19.59 \pm 136.54$ & $121.05 \pm 168.07$ \\
$L_y$ & $39.11 \pm 1092.73$ & $20.72 \pm 471.72$ & $12.52 \pm 556.04$ & $-13.95 \pm 266.13$ & $140.55 \pm 473.93$ & $-255.35 \pm 1174.84$ & $162.22 \pm 1343.78$ & $27.39 \pm 424.14$ & $19.69 \pm 493.20$ \\
$L_z$ & $6.31 \pm 436.14$ & $27.30 \pm 365.15$ & $1062.53 \pm 395.92$ & $97.48 \pm 203.67$ & $737.92 \pm 149.86$ & $-1870.55 \pm 633.08$ & $125.61 \pm 498.32$ & $173.67 \pm 457.48$ & $-757.13 \pm 224.70$ \\
$J_R$ & $778.47 \pm 1013.49$ & $879.40 \pm 671.88$ & $260.21 \pm 210.25$ & $184.54 \pm 100.07$ & $308.60 \pm 146.44$ & $749.41 \pm 421.92$ & $499.10 \pm 282.06$ & $425.27 \pm 234.50$ & $255.10 \pm 142.60$ \\
$J_z$ & $624.20 \pm 383.73$ & $146.64 \pm 136.98$ & $160.84 \pm 137.07$ & $86.01 \pm 53.97$ & $124.57 \pm 85.07$ & $412.75 \pm 258.41$ & $874.51 \pm 274.95$ & $131.56 \pm 115.22$ & $141.48 \pm 112.07$ \\
$E$ & $-155834.24 \pm 29924.96$ & $-156122.06 \pm 17740.35$ & $-163370.99 \pm 11811.41$ & $-201973.11 \pm 12379.59$ & $-171583.66 \pm 9519.88$ & $-127141.10 \pm 11530.99$ & $-149782.72 \pm 12734.69$ & $-173728.59 \pm 14429.00$ & $-172992.10 \pm 10095.83$ \\
$e$ & $0.76 \pm 0.13$ & $0.89 \pm 0.07$ & $0.48 \pm 0.15$ & $0.79 \pm 0.11$ & $0.60 \pm 0.12$ & $0.55 \pm 0.12$ & $0.69 \pm 0.18$ & $0.78 \pm 0.12$ & $0.56 \pm 0.13$ \\
$Z_\mathrm{max}$ & $13.71 \pm 12.62$ & $6.08 \pm 3.99$ & $3.72 \pm 2.25$ & $2.44 \pm 0.99$ & $3.14 \pm 1.41$ & $11.09 \pm 3.56$ & $13.06 \pm 3.41$ & $3.73 \pm 1.93$ & $3.27 \pm 1.61$ \\
$R_\mathrm{peri}$ & $1.80 \pm 1.27$ & $0.76 \pm 0.57$ & $3.55 \pm 1.32$ & $0.64 \pm 0.33$ & $2.25 \pm 0.60$ & $6.26 \pm 1.63$ & $2.78 \pm 2.17$ & $1.17 \pm 0.76$ & $2.44 \pm 0.85$ \\
$R_\mathrm{apo}$ & $16.23 \pm 13.70$ & $14.55 \pm 8.68$ & $10.42 \pm 2.95$ & $5.60 \pm 1.46$ & $9.31 \pm 1.91$ & $22.40 \pm 6.83$ & $14.36 \pm 3.82$ & $9.59 \pm 2.75$ & $8.88 \pm 1.89$ \\
\hline
\end{tabular}
}
\end{table*}

\subsection{FO1: A Compact Metal-Rich Inner-Halo Structure}
\label{subsec:disc_fo1}

FO1 is one of the most statistically robust structures identified in this work, exhibiting the highest bootstrap recovery fraction among all candidate groups ($f_{\rm boot}=82.7\%$; Table~\ref{tab:validation}). Chemically, it is also the most metal-rich candidate population, with $[\mathrm{Fe/H}] = -0.83 \pm 0.08$, placing it substantially above the metallicity range typically associated with the Sequoia structures and the metal-poor accreted halo populations.

Dynamically, FO1 follows a tightly bound prograde orbit ($L_z = 447~\mathrm{kpc~km~s^{-1}}$) with relatively low orbital height ($Z_{\rm max} = 1.83$ kpc) and moderate eccentricity ($e = 0.62 \pm 0.09$). Although its orbital configuration partially overlaps with the inner-halo region occupied by Arjuna-like populations, its comparatively high metallicity and strong clustering stability distinguish it from the broader diffuse halo structures.

The combination of compact orbital structure and enhanced binding energy ($E = -1.95 \times 10^5~\mathrm{km^2~s^{-2}}$) suggests that FO1 may trace the debris of an early relatively massive accretion event that experienced substantial dynamical settling within the inner Milky Way.

\subsection{FO2: The Debris of a Disrupted Globular Cluster}
\label{subsec:disc_fo2}

The most striking chemical anomaly in our sample is found in FO2. As detailed in Table~\ref{tab:chem_new}, all members of this group exhibit a dramatic nitrogen enhancement, with a median [N/Fe] of $+0.83 \pm 0.16$. This exceeds the median [N/Fe] abundance of the recovered GES population by approximately $0.82$ dex and is significantly higher than the nitrogen abundances observed in the other candidate groups.

This extreme [N/Fe] enrichment is coupled with moderately enhanced aluminum ([Al/Fe] = $+0.28 \pm 0.28$) and depleted carbon ([C/Fe] = $-0.36 \pm 0.27$). Crucially, this chemical pattern remains consistent across the group's surface gravity range ($\log g$ from 0.91 to 2.27). Even stars at the higher end of this range display [N/Fe] values exceeding $+1.0$, indicating that the enrichment cannot be explained solely by internal stellar evolutionary processes such as the first dredge-up.

Instead, the combination of high nitrogen, enhanced aluminum, and carbon depletion represents the classic ``second-generation'' abundance pattern commonly associated with stars formed in massive Globular Clusters (GCs).

Furthermore, FO2 fails multiple validation criteria (Table~\ref{tab:validation}), including a negative silhouette score and a low bootstrap recovery fraction ($f_{\rm boot}=16.1\%$). Given that known intact GC members were removed during the initial sample selection, the combination of weak dynamical cohesion and GC-like chemistry strongly suggests that FO2 traces the dispersed debris of a dissolved massive GC, or stars originating from a GC-hosting dwarf galaxy progenitor.

\subsection{FO3: A Dynamically Cold Component in the Inner Halo}
\label{subsec:disc_fo3}

FO3 represents one of the most dynamically unusual structures identified in this work. Despite occupying the low-energy inner halo ($E = -2.01 \times 10^5$ km$^2$ s$^{-2}$), it exhibits a remarkably low eccentricity ($e = 0.38 \pm 0.07$), low radial action ($J_R = 60$ kpc km s$^{-1}$), and near-planar orbital confinement ($Z_{\rm max} = 1.12$ kpc).

Such low-eccentricity configurations are comparatively uncommon among classical accreted halo structures, which are generally dominated by dynamically hot radial orbits. While its strong prograde rotation ($L_z = 596~\mathrm{kpc~km~s^{-1}}$) may superficially resemble disk-like kinematics, its chemistry reveals a more intriguing picture. Remarkably, all six measured chemical dimensions agree with the Helmi Streams reference population within approximately $0.7\sigma$, rendering FO3 chemically indistinguishable from the Helmi population within the published intrinsic scatter. The entire $\chi^2$ excess therefore arises primarily from dynamical dimensions, particularly orbital energy and eccentricity, indicating that FO3 may represent a dynamically evolved or more tightly bound counterpart of an otherwise chemically Helmi-like progenitor.

This orbital configuration is consistent with either a low-inclination accretion event or subsequent orbital circularisation within the evolving Galactic potential. FO3 therefore likely traces an ancient dynamically cold accretion remnant that has become strongly integrated into the inner-halo structure of the Milky Way.

\subsection{FO4: An \texorpdfstring{$\alpha$}{alpha}-Enhanced Inner-Halo Structure}
\label{subsec:disc_fo4}

FO4 exhibits chemical signatures characteristic of accreted stellar populations, although its detailed abundance patterns suggest a distinct evolutionary history relative to the canonical halo merger remnants.

FO4 is notably $\alpha$-enhanced, with [Mg/Fe] = $+0.26 \pm 0.03$ and [Si/Fe] = $+0.27 \pm 0.03$. Relative to the median GES abundance pattern (Table~\ref{tab:chem_known}), these values correspond to enhancements of approximately $+0.72\sigma$ and $+1.06\sigma$, respectively. Although FO4 occupies a similar region of the chemo-dynamical parameter space as GES, its more tightly bound orbit ($E = -1.83 \times 10^5~\mathrm{km^2~s^{-2}}$) and enhanced $\alpha$ abundances suggest that it may represent debris originating from a chemically distinct progenitor or from a more centrally concentrated region of an accreted system.

Its strong cohesion score ($z=-5.76$; Table~\ref{tab:validation}) and stable recovery across multiple UMAP neighbourhood scales further support the interpretation that FO4 represents a physically meaningful inner-halo overdensity rather than a local fluctuation within the broader GES population.

\subsection{FO5: A Low-Aluminum Accreted Remnant Overlapping with Shiva}

\begin{figure*}
    \centering
    \includegraphics[width=\textwidth]{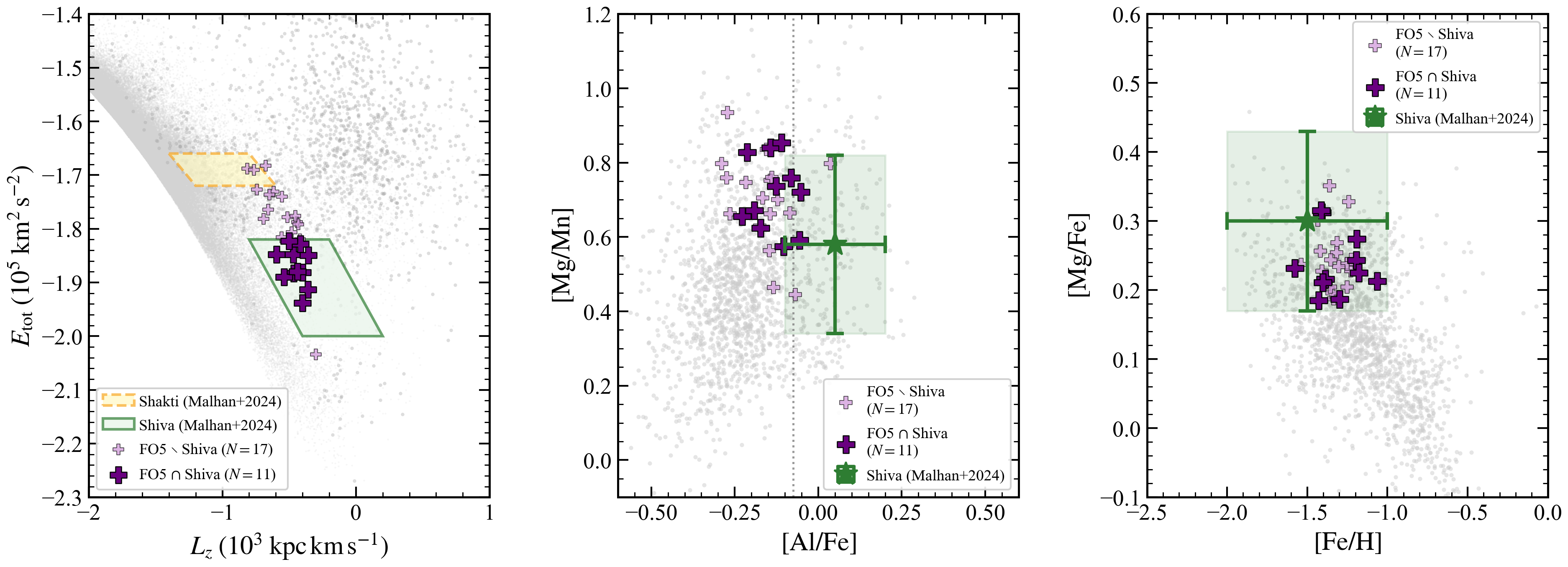}
    \caption{
Comparison between FO5 and the Shiva and Shakti structures proposed by \citet{Malhan2024}. Left: distribution in the $E_{\rm tot}$--$L_z$ plane showing the Shiva (green polygon) and Shakti (orange polygon) selection regions. Dark purple symbols indicate FO5 members located inside the Shiva region ($N=11$), while light purple symbols correspond to FO5 stars outside the Shiva selection ($N=17$). Centre: [Mg/Mn]--[Al/Fe] plane. Right: [Mg/Fe]--[Fe/H] plane. The green star and shaded regions indicate the median abundance properties and dispersions reported for Shiva by \citet{Malhan2024}.}
    \label{fig:shiva_compare}
\end{figure*}

The inner-halo region in the $(E, L_z)$ plane contains substantial overlap between multiple accreted and proto-Galactic populations, complicating the identification of distinct merger remnants. A prominent example is the recently identified Shiva substructure \citep{Malhan2024}. As shown in Figure~\ref{fig:shiva_compare}, our analysis indicates that approximately 39\% of FO5 members ($N=11$) reside within the Shiva selection polygon ($E = -1.8$ to $-2.0 \times 10^5~\mathrm{km^2~s^{-2}}$). However, the detailed chemical characterization of Shiva reveals a fundamental discrepancy with the properties of FO5.

Shiva as a proto-Galactic population characterized by rapid chemical enrichment in a high-density environment, resulting in an Al-rich signature (median $[\mathrm{Al/Fe}] = 0$, with nearly 50\% of stars exhibiting $[\mathrm{Al/Fe}] > 0$). In contrast, FO5 resides within the chemical locus of accreted populations, with a median $[\mathrm{Al/Fe}] = -0.14 \pm 0.08$, consistently below the $-0.075$ dex threshold adopted by \citet{Belokurov2022} to distinguish accreted debris from the ``Aurora'' proto-disk population.

Crucially, this chemical divergence is not driven by averaging over distinct populations. All eleven FO5 members located inside the Shiva polygon individually exhibit negative [Al/Fe] abundances, with a median value of $[\mathrm{Al/Fe}] = -0.15 \pm 0.08$. This star-by-star consistency strongly suggests that even the dynamically overlapping FO5 members belong to the accreted regime rather than to the proto-Galactic population associated with Shiva.

Furthermore, FO5 exhibits a substantial manganese depletion ($[\mathrm{Mn/Fe}] = -0.46 \pm 0.13$). Such low [Mn/Fe] and [Al/Fe] abundances are commonly associated with inefficient chemical enrichment in low-mass progenitor systems \citep{Hasselquist2021}, a formation history that differs from the rapid high-density enrichment scenario proposed for Shiva.

Therefore, although the overlap in $(L_z, E)$ space is striking, it does not necessarily imply a common physical origin. We interpret FO5 as a chemically distinct and robustly accreted remnant that occupies a similar orbital region to Shiva within the dynamically complex inner halo. This result highlights the importance of combining chemical abundance information, particularly [Al/Fe] and [Mn/Fe], with orbital properties in order to disentangle overlapping populations in the inner Milky Way halo.

\subsection{Implications for the Assembly of the Congested Inner Halo}
\label{subsec:disc_synthesis}

\begin{table}
\caption{Mean orbital parameter values for each substructure identified in this work (McMillan 2017 potential). (new candidate groups)}
\label{tab:kin_new}
\resizebox{\columnwidth}{!}{%
\begin{tabular}{lrrrrr}
\hline\hline
\multicolumn{1}{l}{} & \textbf{FO1} & \textbf{FO2} & \textbf{FO3} & \textbf{FO4} & \textbf{FO5} \\
$N$ & 10 & 10 & 15 & 25 & 28 \\
\hline
$v_R$ & $31.55 \pm 58.44$ & $15.73 \pm 93.61$ & $55.09 \pm 60.57$ & $-7.66 \pm 101.06$ & $19.47 \pm 65.45$ \\
$v_\phi$ & $83.18 \pm 24.12$ & $14.09 \pm 85.34$ & $123.69 \pm 24.93$ & $19.72 \pm 45.58$ & $68.23 \pm 14.78$ \\
$v_Z$ & $16.08 \pm 46.06$ & $-16.52 \pm 66.62$ & $20.02 \pm 41.25$ & $-1.25 \pm 59.40$ & $13.81 \pm 48.45$ \\
$L_x$ & $-16.16 \pm 113.83$ & $-0.92 \pm 156.16$ & $-49.91 \pm 123.85$ & $-18.68 \pm 105.30$ & $-65.49 \pm 144.09$ \\
$L_y$ & $-131.32 \pm 231.00$ & $-16.56 \pm 440.93$ & $-118.49 \pm 147.91$ & $6.01 \pm 301.66$ & $-80.04 \pm 343.38$ \\
$L_z$ & $447.02 \pm 93.22$ & $105.48 \pm 427.51$ & $595.52 \pm 119.61$ & $116.19 \pm 317.97$ & $525.11 \pm 133.76$ \\
$J_R$ & $184.75 \pm 99.72$ & $310.87 \pm 85.12$ & $60.22 \pm 23.41$ & $375.52 \pm 179.44$ & $259.15 \pm 107.01$ \\
$J_z$ & $60.26 \pm 30.68$ & $159.88 \pm 123.57$ & $47.18 \pm 32.12$ & $74.66 \pm 55.63$ & $106.77 \pm 61.89$ \\
$E$ & $-194939.89 \pm 10444.67$ & $-178517.77 \pm 6907.01$ & $-201486.53 \pm 6808.63$ & $-182666.14 \pm 9448.18$ & $-181195.10 \pm 8127.97$ \\
$e$ & $0.62 \pm 0.09$ & $0.77 \pm 0.14$ & $0.38 \pm 0.07$ & $0.80 \pm 0.08$ & $0.68 \pm 0.05$ \\
$Z_\mathrm{max}$ & $1.83 \pm 0.80$ & $3.68 \pm 1.93$ & $1.12 \pm 0.49$ & $2.84 \pm 0.98$ & $2.73 \pm 0.96$ \\
$R_\mathrm{peri}$ & $1.40 \pm 0.20$ & $1.09 \pm 0.66$ & $2.15 \pm 0.45$ & $0.89 \pm 0.40$ & $1.51 \pm 0.30$ \\
$R_\mathrm{apo}$ & $6.12 \pm 1.34$ & $8.41 \pm 1.23$ & $4.78 \pm 0.58$ & $8.04 \pm 1.60$ & $7.87 \pm 1.37$ \\
\hline
\end{tabular}
}
\end{table}

The collective properties of our newly identified structures (FO1--FO5) provide critical insights into both the assembly history of the inner Milky Way and the methodological challenges of Galactic archaeology. Traditionally, the identification of halo substructures has relied heavily on integrals-of-motion (IoM) space, particularly the $E_{\rm tot}$--$L_z$ plane \citep{Koppelman2019, Naidu2020}. However, our findings demonstrate that the low-energy inner halo ($E_{\rm tot} \leq -1.8 \times 10^5~\mathrm{km^2~s^{-2}}$) is a highly congested regime in which purely dynamical clustering becomes insufficient for reliably disentangling physically distinct populations.

The clearest evidence for this complexity arises from the contrasting behaviour of FO3 and FO5. As discussed in Section~\ref{subsec:disc_fo3}, FO3 is chemically consistent with the Helmi Streams while remaining dynamically distinct. Conversely, FO5 occupies a similar dynamical region to the proto-Galactic Shiva substructure while exhibiting clearly distinct chemical abundance patterns. If our analysis had relied solely on orbital selections, FO5 would likely have been absorbed into the Shiva population despite its chemically accreted nature. These degeneracies underscore the fundamental importance of high-dimensional chemo-dynamical clustering: while orbital properties evolve and phase-mix over cosmic time, stellar chemical abundances remain largely unaffected by this dynamical evolution and preserve the nucleosynthetic signatures of their birth environments \citep{Freeman2002, Ting2015}.

Beyond major dwarf-galaxy merger remnants, the physical nature of these highly bound structures highlights the diversity of accretion and disruption channels that contributed to the formation of the early proto-halo. A prominent example is FO2, whose extreme nitrogen enhancement and carbon depletion introduce a fundamentally distinct physical channel into our synthesis. Rather than originating from a classical dwarf galaxy accretion event, FO2 strongly traces the dispersed debris of a dissolved massive Globular Cluster (GC) or a GC-hosting progenitor system, emphasizing that star-cluster disruption contributes significantly even within the most dynamically congested regions of the inner halo.

\begin{table*}
\centering
\caption{%
  Validation metrics for the five new candidate substructures (FO1--FO5).
  Columns report: number of members ($N$);
  combined chi-square distance to the closest literature reference from
  \citet{Horta2023} ($\chi^{2}_{\rm comb}$);
  cohesion $z$-score and permutation $p$-value from the
  12-dimensional compactness test ($N_{\rm perm}=1000$);
  mean UMAP silhouette coefficient ($\langle s_i \rangle$);
  bootstrap co-clustering percentage
  ($f_{\rm boot}$; $N_{\rm boot}=30$ full-pipeline resamples);
  and UMAP co-clustering fractions ($f_{\rm co}$)
  measured relative to the reference embedding
  ($n_{\rm neighbors}=20$) for alternative neighbourhood scales.%
}
\label{tab:validation}
\setlength{\tabcolsep}{6pt}
\begin{tabular}{l r r r r r r r r r l}
\toprule
\multirow{2}{*}{Group}
  & \multirow{2}{*}{$N$}
  & \multirow{2}{*}{$\chi^{2}_{\rm comb}$}
  & \multicolumn{2}{c}{Cohesion (12D)}
  & \multirow{2}{*}{$\langle s_i \rangle$}
  & \multirow{2}{*}{$f_{\rm boot}$ (\%)}
  & \multicolumn{3}{c}{UMAP $f_{\rm co}$}
  & \multirow{2}{*}{Status} \\
\cmidrule(lr){4-5}
\cmidrule(lr){8-10}
  & & &
  $z$ & $p$ &
  & &
  $n_{\rm nb}=10$ &
  $n_{\rm nb}=30$ &
  $n_{\rm nb}=50$ & \\
\midrule
FO1 & 10 & 3.162 & $-4.23$ & $<0.001$ & $\phantom{-}0.521$ & 82.7 & 1.00 & 1.00 & 1.00 & Robust     \\
FO2 & 10 & 2.138 & $-1.12$ & $0.091$  & $-0.025$           & 16.1 & 1.00 & 0.00 & 0.00 & Tentative  \\
FO3 & 15 & 3.416 & $-4.52$ & $<0.001$ & $\phantom{-}0.317$ & 71.6 & 1.00 & 0.87 & 0.87 & Robust     \\
FO4 & 25 & 2.171 & $-5.76$ & $<0.001$ & $\phantom{-}0.261$ & 50.9 & 0.88 & 0.84 & 0.84 & Robust     \\
FO5 & 28 & 2.156 & $-6.12$ & $<0.001$ & $\phantom{-}0.268$ & 37.3 & 0.86 & 0.75 & 0.96 & Robust     \\
\bottomrule
\end{tabular}
\begin{tablenotes}
  \small
  \item FO2 fails three of the four validation criteria
    (cohesion $p=0.091$, $\langle s_i \rangle < 0$,
    and $f_{\rm co}=0.00$ for $n_{\rm neighbors}\geq30$)
    and is therefore classified as a tentative candidate pending
    confirmation with larger samples.
  \item The cohesion $z$-score measures the compactness of each group
    relative to random equal-sized draws from the ex-situ parent sample.
    $f_{\rm boot}$ represents the fraction of bootstrap realisations
    in which the majority of members remain clustered together.
    $f_{\rm co}$ denotes the UMAP co-clustering fraction relative
    to the reference embedding.
\end{tablenotes}
\end{table*}

The remaining accreted structures exhibit varying degrees of dynamical evolution within the Galactic potential, reflecting diverse infall conditions rather than a uniform orbital history. FO3 displays a remarkably low eccentricity ($e = 0.38$), consistent with a dynamically cold accreted component that may have experienced substantial orbital circularisation or been accreted along a low-eccentricity trajectory. In contrast, FO1 represents a strongly settled yet tightly bound inner-halo structure with a substantially higher eccentricity ($e = 0.62$). This wide diversity of orbital properties demonstrates that even among structures deposited onto low-energy inner halo, the dynamical histories of accreted systems remain highly heterogeneous.

The challenge of associating such deeply bound debris with canonical merger remnants is further complicated by the presence of intrinsic chemical gradients within progenitor galaxies. As discussed by \citet{Amarante2022}, spatial variations in metallicity and abundance ratios within a dwarf galaxy progenitor can naturally map onto the present-day chemo-dynamical distribution of its disrupted debris, since material originating from different regions of the satellite is stripped at different stages of the merger process. Consequently, the distinct chemical signatures observed in structures such as FO4 and FO5 --- including their enhanced $\alpha$ abundances and depleted manganese values --- may reflect internal nucleosynthetic variations within their progenitor systems rather than entirely separate merger events.

Together, these systems highlight the diversity of progenitor populations contributing to the formation of the inner Milky Way halo and demonstrate that the innermost halo is unlikely to be dominated by a single major merger event. Instead, it appears to preserve the chemically distinguishable debris of several ancient accretion channels, including massive dwarf-galaxy mergers, chemically inefficient low-mass systems, and disrupted star clusters.

Further progress will require substantially larger samples of chemically characterized stars in the low-energy inner halo. Expanding the membership of these candidate structures through forthcoming data releases from large spectroscopic surveys such as \textsl{SDSS-V} and 4MOST, together with future extensions of \textsl{Gaia} radial velocity measurements, will be critical for testing the physical reality of these groups and constraining the masses, star-formation histories, and chemical evolution pathways of their progenitor systems.

\section{Conclusions}
\label{sec:conclusions}

The inner Galactic halo retains a rich archaeological record of ancient accretion events; however, its pronounced chemical and dynamical complexity has historically resisted systematic decomposition. In this work, we have shown that a high-dimensional chemo-dynamical analysis of the \textsl{SDSS-V MWM} DR19 and \textsl{Gaia} DR3 datasets can successfully disentangle these overlapping field populations in a purely data-driven manner, without imposing kinematic pre-selection criteria.

Our main findings are summarized as follows:

\begin{itemize}

\item \textbf{Recovery of Canonical Substructures:}
We successfully recovered nine kinematic groupings corresponding to seven distinct previously known halo substructures, including major merger remnants such as \textsl{Gaia}-Enceladus/Sausage (GES), the Helmi Streams, Thamnos, Heracles, LMS-1, Arjuna and three independently identified kinematic variants of Sequoia \citep{Koppelman2019, Myeong2019, Naidu2020}. The chemo-dynamical properties of these recovered groups closely match the reference values reported in the literature, validating the reliability of our unsupervised framework.

\item \textbf{Discovery of New Inner-Halo Overdensities:}
We identified five new candidate substructures, designated FO1--FO5, concentrated in the low-energy inner halo ($E_{\rm tot} < -1.8 \times 10^5~\mathrm{km^2~s^{-2}}$). Through statistical validation---including permutation-based 12D cohesion tests, UMAP silhouette analysis, and bootstrap resampling---we demonstrated that FO1, FO3, FO4, and FO5 represent robust chemo-dynamical overdensities. Furthermore, a cross-match with the recent catalogue of \citet{Dodd2025} confirmed that these structures do not overlap with known local halo streams.

\item \textbf{Diverse Progenitor Origins and Star-Cluster Disruption:}
The detailed chemical profiles of our candidates point toward a heterogeneous assembly history. While the enhanced $\alpha$ abundances and low manganese content observed in FO4 and FO5 likely reflect different chemical enrichment efficiencies within dwarf galaxy progenitors, FO2 traces a fundamentally different physical channel. Although FO2 fails our statistical cohesion criterion ($p = 0.091 > 0.05$), which marks it as a tentative candidate, it exhibits a striking chemical profile characterized by extreme nitrogen enhancement ($[\mathrm{N/Fe}] = +0.83 \pm 0.16$), moderate aluminum enhancement, and carbon depletion. This distinct pattern provides strong evidence that FO2 traces the tidal debris of a dissolved massive Globular Cluster (GC) or a GC-hosting system.

\item \textbf{The Dynamically Crowded Nature of the Inner Halo:}
Our results highlight the critical role of high-dimensional chemical information in resolving degeneracies caused by orbital overlap in the inner halo. We find that FO3 is chemically nearly indistinguishable from the Helmi Streams while remaining dynamically much colder ($e = 0.38$). Conversely, FO5 shows the reverse pattern: it overlaps dynamically with the proto-Galactic Shiva structure yet is chemically distinct from it, betraying an independent accreted origin. These findings demonstrate that purely dynamical classifications in crowded regimes can lead to significant misidentifications.

\end{itemize}

We note that some of our newly discovered candidate groups (such as FO1 and FO2) currently rely on small sample sizes ($N=10$), meaning their full spatial extents and detailed chemical gradients remain unconstrained. Nevertheless, the discovery of these chemically and dynamically distinct structures suggests that the inner halo preserves evidence of multiple ancient accretion channels. In particular, constraining the progenitor mass function of the accreted satellites responsible for these structures, and quantifying the contribution of dissolved GCs to the inner-halo field population, will be key goals for the next generation of spectroscopic surveys. Future data releases from \textsl{SDSS-V}, 4MOST, and expanded \textsl{Gaia} radial velocity catalogues will be crucial for enlarging the membership of these structures and further unravelling the complex assembly history of the Galactic inner halo. Our results suggest that the innermost regions of the Galactic halo remain far from fully mapped and likely harbours additional chemically distinct relics of the Milky Way's earliest assembly history.

\begin{acknowledgments}
This work was supported by the Scientific and Technological Research Council of Turkey (TÜBİTAK) under project code MFAG-123F227, MFAG-125F465 and program 2211-C. This study was funded by the Scientific Research Projects Coordination Unit of the Istanbul University. Project numbers FBA-2023-39380, and FDK-2025-41537.
\end{acknowledgments}

\appendix
\renewcommand{\thesection}{\Alph{section}}
\renewcommand{\thefigure}{A\arabic{figure}}
\setcounter{figure}{0}

\section{Chemical Abundance Calibrations}
\label{sec:app_correction}

In this appendix, we provide the comprehensive diagnostic visualisations for the empirical surface gravity ($\log~g$) calibration described in Section~\ref{sec:correction}. Due to the high dimensionality of the chemical parameter space and the large size of the diagnostic plots, the complete distribution patterns are compiled here to avoid disrupting the flow of the primary text. 

\begin{figure}
    \includegraphics[width=0.5\textwidth]{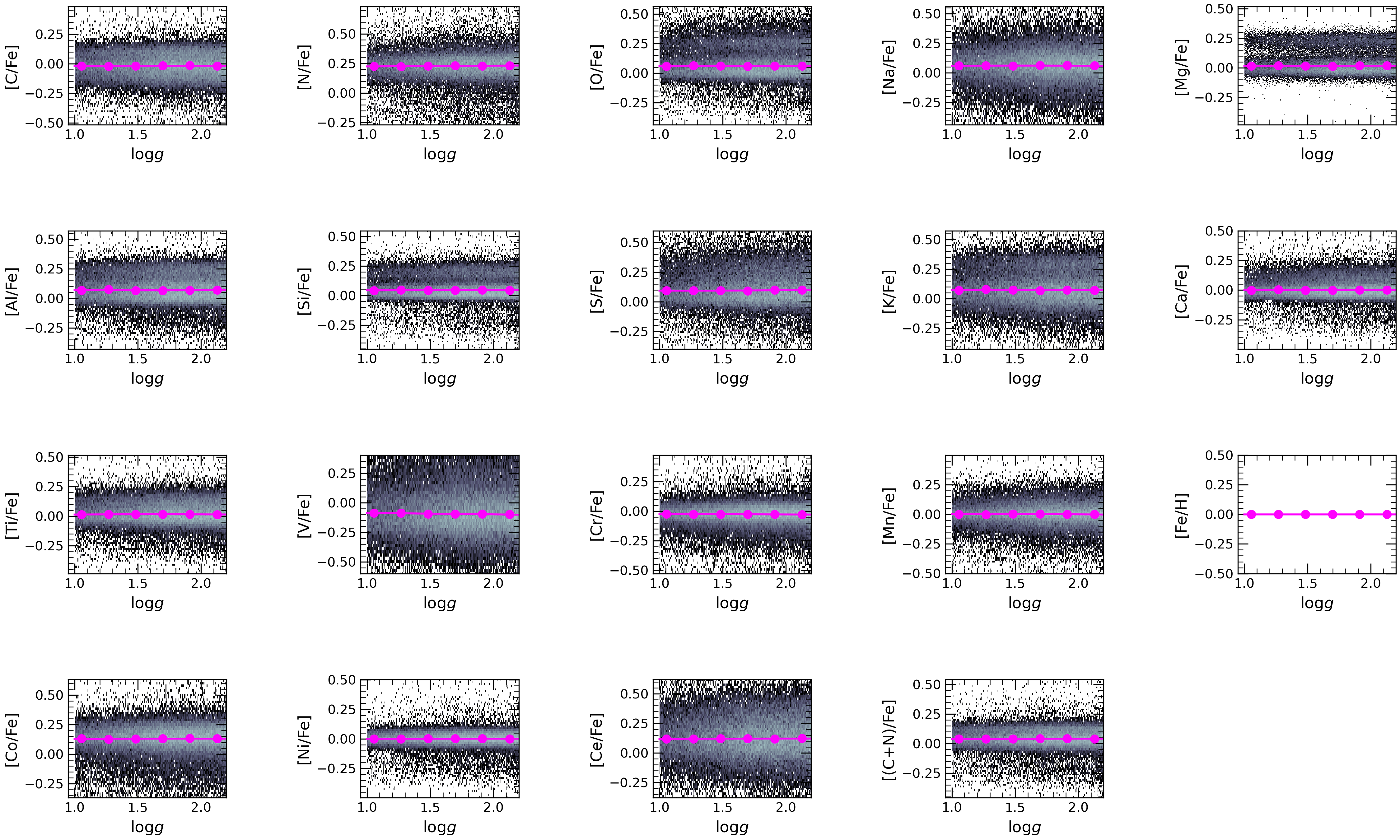}
    \includegraphics[width=0.5\textwidth]{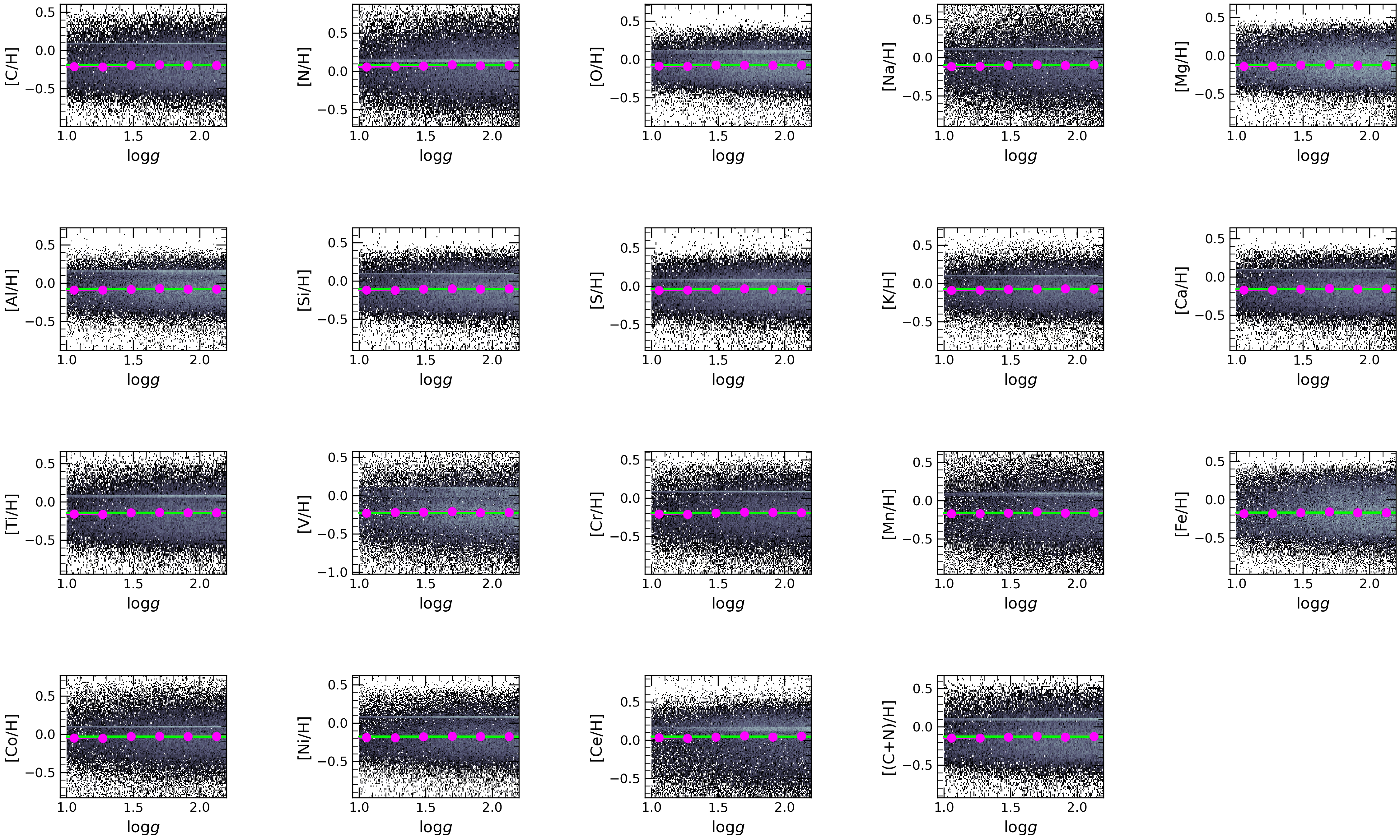}
    \includegraphics[width=0.5\textwidth]{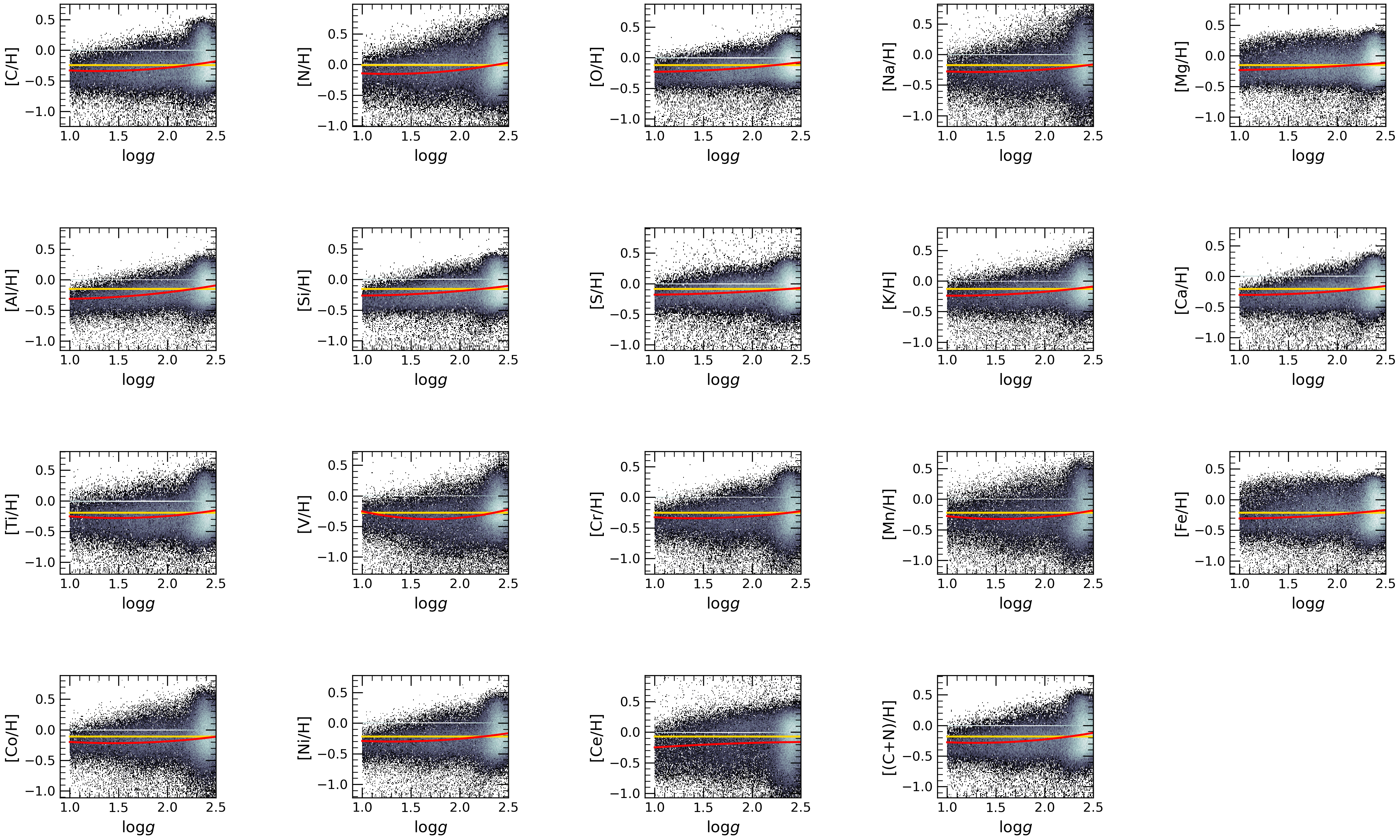}
    \caption{
Validation of the abundance corrections applied to the high-$\alpha$ subsample. Top panel: Stellar abundance ratios [X/H] as a function of surface gravity $\log g$ prior to the correction procedure. The two-dimensional density distribution and running medians reveal systematic $\log g$-dependent trends introduced by the APOGEE abundance pipeline. Middle panel: Same as the top panel after applying the abundance correction procedure. The reduced slope of the running medians demonstrates that the systematic trends are significantly suppressed. Bottom panel: Corrected abundance ratios [X/Fe] as a function of $\log g$. The nearly flat running medians indicate that the corrected abundance ratios are largely free from residual surface-gravity-dependent systematics, validating their use in the subsequent chemo-dynamical analysis.}
    \label{fig:abundance_correction}
\end{figure}

Figure~\ref{fig:abundance_correction} displays the density distributions of the measured chemical abundance ratios as a function of $\log~g$ both before (raw pipeline values) and after the application of our second-order polynomial correction. As illustrated, the systematic tilts and non-linear parameter dependencies present in the uncalibrated data are successfully modernised and flattened. This correction ensures that the chemical tagging and high-dimensional clustering analysis performed in the main body of this work are driven strictly by nucleosynthetic origins rather than systematic offsets tied to the evolutionary stages of the sample stars.

\software{astropy \citep{Astropy2013, Astropy2018, Astropy2022},
          galpy \citep{galpy}, python \citep{Python}, matplotlib \citep{Matplotlib}, scikit-learn \citep{scikit-learn}
          }




\bibliography{references}{}
\bibliographystyle{aasjournalv7}



\end{document}